\documentclass[a4paper,11pt]{article}

\usepackage{jheppub_mod} 
\usepackage[T1]{fontenc} 

\usepackage{hyperref}
\usepackage{graphicx}
\usepackage{amsmath,amssymb,slashed}
\usepackage{booktabs,tabulary}
\usepackage{dsfont}
\usepackage{color}
\usepackage[dvipsnames]{xcolor}
\usepackage{multirow}
\usepackage{subcaption}
\usepackage{tikz-feynman}
\usepackage[compat=1.1.0]{tikz-feynhand}
\usetikzlibrary{decorations.markings,arrows}
\usetikzlibrary{arrows.meta}
\usepackage{extarrows} 
\usepackage[normalem]{ulem}

\newcommand{\mpip}{M_{\pi^\pm}}
\newcommand{\mpi}{M_{\pi}}
\newcommand{\mpii}{M_{\pi^0}}
\newcommand{\meta}{M_{\eta}}
\newcommand{\metap}{M_{\eta'}}

\newcommand{\kappaeta}{\kappa\etapi}

\newcommand{\MeV}{\,\text{MeV}}

\newcommand{\beq}{\begin{equation}}
\newcommand{\eeq}{\end{equation}}
\newcommand{\diff}{\text{d}}

\newcommand{\sth}{s_\text{th}}
\newcommand{\tth}{t_\text{th}}

\newcommand{\M}{\mathcal{M}}
\newcommand{\N}{\mathcal{N}}

\newcommand{\A}{\mathcal{A}}

\newcommand{\G}{\mathcal{G}}
\renewcommand{\L}{\mathcal{L}}
\renewcommand{\H}{\mathcal{H}}
\newcommand{\B}{\mathcal{B}}
\newcommand{\disc}{\text{disc}\,}
\renewcommand{\Re}{\text{Re}\,}

\newcommand{\diag}{\text{diag}}
\newcommand{\eqand}{\hspace{.5cm} \mathrm{and} \hspace{.5cm}}
\newcommand{\eqor}{\hspace{.5cm} \mathrm{or} \hspace{.5cm}}
\newcommand{\eqwith}{\, ,\hspace{.5cm} \mathrm{with} \hspace{.5cm}}
\newcommand{\pipi}{_{\pi\pi}}

\newcommand{\fI}{f^\text{I}}

\newcommand{\etapeta}{_{\eta'\eta}}
\newcommand{\etapin}{_{\eta\pi}}
\newcommand{\etapi}{_{\eta\pi}}
\newcommand{\munu}{{\mu\nu}}
\def\vev#1{\big\langle #1 \big\rangle}
\newcommand{\hc}{\text{h.c.}}

\renewcommand{\O}{\mathcal{O}} 
\newcommand{\Fpivec}{F_{\pi}^V}
\newcommand{\Fpiveccon}{F_{\pi}^{V\ast}}

\makeatletter
\DeclareRobustCommand{\cev}[1]{%
  {\mathpalette\do@cev{#1}}%
}
\newcommand{\do@cev}[2]{%
  \vbox{\offinterlineskip
    \sbox\z@{$\m@th#1 x$}%
    \ialign{##\cr
      \hidewidth\reflectbox{$\m@th#1\vec{}\mkern4mu$}\hidewidth\cr
      \noalign{\kern-\ht\z@}
      $\m@th#1#2$\cr
    }%
  }%
}
\makeatother

\def\vecsign{\mathchar"017E}
\def\dvecsign{\smash{\stackon[-2.17pt]{\vecsign}{\rotatebox{180}{$\vecsign$}}}}
\def\dvec#1{\def\useanchorwidth{T}\stackon[-4.2pt]{#1}{\,\dvecsign}}
\usepackage{stackengine}
\stackMath

\tikzfeynmanset{
  fermion1/.style={
    /tikz/postaction={
      /tikz/decoration={
        markings,
        mark=at position 0.6 with {
        \arrow[scale=1.2,>=stealth]{>};
        },
      },
      /tikz/decorate=true,
    },
  },
}

\makeatletter
	\newenvironment{aligneq}
    {\begin{equation}\begin{alignedat}{20}}{\end{alignedat}\end{equation}}
    \usepackage{etoolbox}
    \newcommand*\NoIndentAfterEnv[1]{%
  \AfterEndEnvironment{#1}{\par\@afterindentfalse\@afterheading}}
\makeatother
\NoIndentAfterEnv{aligneq}

\usetikzlibrary{calc}
\makeatletter
\newcommand{\boxwidth}{0.7pt}
\newcommand{\RedBox}[1]{\textcolor{Red}{%
\tikz[baseline={([yshift=-1ex]current bounding box.center)}] \node [rectangle, minimum width=1ex,rounded corners,draw, line width=\boxwidth] {\normalcolor\m@th$\displaystyle#1$};}}
\newcommand{\BlueBox}[1]{\textcolor{blue}{%
\tikz[baseline={([yshift=-1ex]current bounding box.center)}] \node [rectangle, minimum width=1ex,rounded corners,draw, line width=\boxwidth] {\normalcolor\m@th$\displaystyle#1$};}}
\newcommand{\GreenBox}[1]{\textcolor{Green}{%
\tikz[baseline={([yshift=-1ex]current bounding box.center)}] \node [rectangle, minimum width=1ex,rounded corners,draw, line width=\boxwidth] {\normalcolor\m@th$\displaystyle#1$};}}
 \makeatother

\allowdisplaybreaks[1]

\title{\boldmath{Correlations of $C$ and $CP$ violation in $\eta\to \pi^0\ell^+\ell^-$ and $\eta'\to \eta\ell^+\ell^-$}}

\author[a]{Hakan Akdag,}
\author[a]{Bastian Kubis,}
\author[b]{and Andreas Wirzba}

\affiliation[a]{
Helmholtz-Institut f\"ur Strahlen- und Kernphysik (Theorie) and \\
Bethe Center for Theoretical Physics, Universit\"at Bonn, 53115 Bonn, Germany}
\affiliation[b]{Institut  f\"{u}r Kernphysik (Theorie), 
           Institute for Advanced Simulation, and \\
           J\"ulich Center for Hadron Physics,
           Forschungszentrum J\"ulich,  
           52425 J\"{u}lich, Germany}

\emailAdd{akdag@hiskp.uni-bonn.de}
\emailAdd{kubis@hiskp.uni-bonn.de}
\emailAdd{a.wirzba@fz-juelich.de}

\abstract
{Based on recent progress in the systematic analysis of $C$ and $CP$ violation in the light-meson sector, we calculate the $C$-odd transition amplitudes $\eta\to\pi^0\ell^+\ell^-$ and $\eta'\to\eta\ell^+\ell^-$.
Focusing on long-distance contributions driven by the lowest-lying hadronic intermediate states, we work out the correlations between these beyond-the-Standard-Model signals and the Dalitz-plot asymmetries in $\eta \rightarrow \pi^0 \pi^+ \pi^-$ and $\eta' \rightarrow \eta \pi^+ \pi^- $, using dispersion theory. 
}

\begin{document} 
\maketitle

\section{Introduction}
\label{sec:intro}

In the recent past new light was shed on analyzing patterns of $C$- and $CP$-odd signals in hadronic decays concerning the $\eta$-sector. First, Ref.~\cite{Gardner:2019nid} provided a new theoretical framework for $C$~violation in $\eta\to\pi^0\pi^+\pi^-$, which was neglected since the 1960s and allows one to decompose the decay amplitude into contributions of different isospin transitions, hence to disentangle the underlying beyond-the-Standard-Model (BSM) operators of isospin $I=0$ and $I=2$. This analysis has been improved and extended to $\eta'\to\pi^0\pi^+\pi^-$ as well as $\eta'\to\eta\pi^+\pi^-$ in Refs.~\cite{Akdag:2021efj,Akdag:2023oob}, which relied on the dispersion-theoretical Khuri--Treiman framework. Motivated by these analyses, Ref.~\cite{Akdag:2022sbn} derived the full set of $C$- and $CP$-violating quark-level operators for flavor-conserving, lepton- and baryon-number-preserving transitions in the low-energy effective field theory (LEFT) and matched them onto a $T$-odd and $P$-even (ToPe) analog of chiral perturbation theory (ToPe$\chi$PT) (cf.\ also Refs.~\cite{Shi:2017ffh,Gardner:prep}). The latter allows us to access all corresponding $C$- and $CP$-odd mesonic interactions. 

One may use the formalisms addressed in the previous paragraph to investigate the correlation of $T$-odd and $P$-even forces between \emph{different} decays. In this sense we consider $C$- and $CP$-violating radiative decays of $\eta$ and $\eta'$. In the following, we denote both mesons with $\eta^{(\prime)}$. Given that $\eta^{(\prime)}$ as well as $\pi^0$ have the eigenvalue $C=+1$ but photons have $C=-1$, $C$ is violated in general if $\eta^{(\prime)}$ decays into an arbitrary number of uncharged pions and an \emph{odd} number of photons. This consideration also holds for radiative decays of the $\eta'$ into an $\eta$.  In the following we will focus on the radiative $C$-odd decays $\eta\to \pi^0\gamma^{(\ast)}$ and $\eta'\to \eta\gamma^{(\ast)}$. To shorten the notation we will refer to both processes collectively by $X\to Y\gamma^{(\ast)}$. Angular momentum conservation demands the final state to be in a relative $P$-wave. Consequently, parity is conserved and the decays at hand additionally violate $CP$, thus offering an opportunity to investigate ToPe forces. 
The decay into a real, transverse photon violates both gauge invariance and the conservation of angular momentum~\cite{Sakurai:1964, Gan:2020aco}. Therefore, the focus shall be laid on $X\rightarrow Y \gamma^\ast \to Y \ell^+\ell^- $, where the off-shell photon decays subsequently into a pair of charged leptons. 
At the theoretical front, the investigation of this BSM one-photon exchange urgently requires an update~\cite{Bernstein:1965hj,Barrett:1965ia} in comparison to analyses of the SM contribution, cf.\ Refs.~\cite{LlewellynSmith:1967, Cheng:1967zza, Smith:1968, Ng:1992, Ng:1993, Escribano:2020rfs, Schafer:2023qtl}, as well as studies of other BSM effects in these decays~\cite{Gan:2020aco, Escribano:2022zgm}. From an experimental point of view, bounds on all the leptonic channels have already been set~\cite{Dzhelyadin:1980ti, CLEO:1999nsy, WASA-at-COSY:2018jdv} and may become more stringent in future measurements~\cite{Gan:2009zzd, JEF:2014, Gan:2015nyc, JEF:2017, Gatto:2016rae, Gatto:2019dhj, REDTOP:2022slw}.

\begin{table}
\centering
\renewcommand{\arraystretch}{1.3}
\begin{tabular}{lcc}
\toprule
branching ratio & SM prediction~\cite{Schafer:2023qtl} & experimental limit \\
\midrule
$\B(\eta\to\pi^0 e^+e^-)$& $1.36(15)\times 10^{-9}$ & $< 7.5 \times 10^{-6}$~\cite{WASA-at-COSY:2018jdv}\\
$\B(\eta\to\pi^0 \mu^+\mu^-)$& $0.67(7)\times 10^{-9}$ & $< 5 \times 10^{-6}$~\cite{Dzhelyadin:1980ti}\\
$\B(\eta'\to\pi^0 e^+e^-)$& $3.30(28)\times 10^{-9}$ & $< 1.4 \times 10^{-3}$~\cite{CLEO:1999nsy}\\
$\B(\eta'\to\pi^0 \mu^+\mu^-)$& $1.81(16)\times 10^{-9}$ & $< 6 \times 10^{-5}$~\cite{Dzhelyadin:1980ti}\\
$\B(\eta'\to\eta e^+e^-)$& $0.50(4)\times 10^{-9}$ & $< 2.4 \times 10^{-3}$~\cite{CLEO:1999nsy}\\
$\B(\eta'\to\eta \mu^+\mu^-)$& $0.240(21)\times 10^{-9}$ & $< 1.5 \times 10^{-5}$~\cite{Dzhelyadin:1980ti}\\
\bottomrule
\end{tabular}
\renewcommand{\arraystretch}{1.0}
\caption{
Standard-Model predictions for dilepton branching ratios based on the $C$-even two-photon mechanism~\cite{Schafer:2023qtl}, as well as corresponding experimental upper limits. Note that below only the cases $\eta\to\pi^0\ell^+\ell^-$ and $\eta'\to\eta \ell^+\ell^-$ will be  considered further.
\label{tab:BR+SM}}
\end{table}

We summarize the currently most stringent experimental upper bounds on the $X\to Y\ell^+\ell^-$ branching ratios in Table~\ref{tab:BR+SM}, and contrast them with the corresponding SM predictions based on the $C$-conserving two-photon mechanism~\cite{Schafer:2023qtl}. 
We observe that those predictions are below the current limits by large factors, between $5\times 10^3$ (for $\eta\to\pi^0 e^+e^-$) and $5\times10^6$ (for $\eta'\to\eta e^+e^-$).  
We will therefore perform the analysis in the following in the spirit that we assume the SM contribution to be small, and any observable or observed signal to be a sign of a $C$-odd single-photon mechanism.\footnote{Similarly, we disregard other potential BSM effects, such as of $CP$-even weakly coupled scalars; see, e.g., Ref.~\cite{Gan:2020aco} and references therein.}  Note furthermore that $C$-even and $C$-odd amplitudes cannot interfere on the level of the branching ratio; such interference effects could only induce Dalitz-plot asymmetries (cf.\ Refs.~\cite{Gardner:2019nid,Akdag:2021efj,Akdag:2023oob}), which would only be sizable if both amplitudes are of comparable magnitude.  We therefore refrain from discussing such interference effects in any detail.

We emphasize the special role of $CP$~violation in flavor-\emph{conserving} transitions such as all $\eta$ and $\eta'$ decays discussed here.  In contrast to similar kaon decays $K\to\pi\ell^+\ell^-$~\cite{DAmbrosio:1996lam, DAmbrosio:1998gur}, SM $CP$-odd contributions induced by the weak interactions are very strongly suppressed, as any weak phase cancels at one loop, and any such contribution only depends on the squared moduli of the Cabibbo--Kobayashi--Maskawa (CKM) matrix elements.  This is not unlike the situation for electric dipole moments of nucleons, which are considered quasi-free of any CKM-matrix-induced SM background~\cite{Shabalin:1978rs,Khriplovich:1985jr,Czarnecki:1997bu,Pospelov:2005pr}.

Assuming that the underlying new physics generating the $C$- and $CP$-odd decays $X\to Y\ell^+\ell^-$ originates from sources at some high-energy scale $\Lambda$, there are in principle three dominant mechanisms to consider:
\begin{enumerate}
\item short-distance contributions to the dilepton final state,
\item long-distance contributions caused by $C$- and $CP$-odd photon--hadron couplings,
\item long-distance contributions induced by hadronic intermediate states.
\end{enumerate}
For the first two classes we rely on ToPe$\chi$PT as proposed in Ref.~\cite{Akdag:2022sbn}. 
One intricacy of the contribution by hadronic intermediate states is that the subsequent photon is allowed to have both isoscalar and isovector components.
To predict the involved isovector transitions in a model-independent way, one can utilize the  $X\to Y\pi^+\pi^-$ amplitudes derived non-perturbatively in the Khuri--Treiman framework~\cite{Akdag:2021efj,Akdag:2023oob} and establish dispersion relations for the respective transition form factors.\footnote{This relation is the reason why we largely ignore $\eta'\to\pi^0\ell^+\ell^-$ in this article: the corresponding hadronic decay $\eta'\to\pi^0\pi^+\pi^-$ is comparatively rare, such that no useful limits on $C$~violation therein can be derived so far~\cite{BESIII:2016tdb,Akdag:2021efj,Akdag:2023oob}.}
Analogous relations have previously been derived for the decays $\omega,\,\phi,\,J/\psi\to \pi^0\gamma^\ast$~\cite{Koepp:1974da, Schneider:2012ez, Danilkin:2014cra, Kubis:2014gka, JPAC:2020umo}, which are compatible with conservation of all discrete symmetries. 
In addition, we sketch an idea of how to evaluate the isoscalar contribution of the photon, employing a less sophisticated, but still symmetry-driven, vector-meson-dominance (VMD) model for the decay $X\to Y\gamma^*$. By an analytic continuation of the three-body amplitudes $X \to Y \pi^+\pi^-$ to the second Riemann sheet we can extract $\rho YX$ couplings, which can be related to the relevant ones with the same total isospin in the VMD model using $SU(3)$ symmetry and naive dimensional analysis (NDA).

To extract observables of the $C$- and $CP$-violating contribution in $X\to Y\ell^+\ell^-$ driven by a one-photon exchange we pursue the following strategy. 
First, we consider the phenomenology behind the three mechanisms mentioned above in Sect.~\ref{sec:phenomenology}. For this purpose we lay out the basic definitions of kinematics and relate the amplitude to the (differential) decay widths in Sect.~\ref{sec:kinematics}. We discuss the short-range semi-leptonic operators, the long-range direct photon--hadron couplings, and the long-range hadronic contributions on a general level in Sects.~\ref{sec:Direct_Semi-Leptonic}--\ref{sec:hadronic_long_range}, respectively. Subsequently, Sect.~\ref{sec:discussion_phenomenology} includes a discussion of the feasibility of these contributions. The remainder of the article solely focuses on long-distance contributions with hadronic intermediate states.
In Sect.~\ref{sec:Hadronic_long_range_isovector} we investigate the isovector contributions to these hadronic long-range effects. For this purpose, we first sketch the $C$- and $CP$-odd contributions to the decays $X\to Y\pi^+\pi^-$ in  Sect.~\ref{sec:KT_Eta}, which serve as input to the respective transition form factors. The computation of the latter is discussed in Sect.~\ref{sec:Formfac_computation}. 
In Sect.~\ref{sec:analytic_cont.}, we extract the corresponding $C$-odd couplings of the $\rho(770)$ resonance to $\eta\pi^0$ and $\eta'\eta$ by analytic continuation in the complex-energy plane. 
Subsequently, we estimate the size of the hadronic long-range effects in the isoscalar parts in Sect.~\ref{sec:Hadronic_long_range_isoscalar}.
Finally, we present the predicted upper limits on the branching ratios in Sect.~\ref{sec:results} and close with a short summary and outlook in Sect.~\ref{sec:summary}.

\section{Phenomenology}
\label{sec:phenomenology}
\begin{figure}[t!]
\centering
		\[
		\resizebox{0.965\textwidth}{!}{
		\begin{tikzpicture}
		\begin{feynman}[large]
		\begin{feynhand}[baseline=(a)]
		\vertex (a);
		\vertex [left=-5cm of a] (xx);
		\vertex [left=0.0cm of a] (a1){\(X\)};
		\vertex [right=1.5of a] (b);
		\vertex [right=1.4of b] (e);
		\vertex [right=0.1of e] (e1){\(Y\)};
		\vertex [above=.8of e] (eup){\( \ell^+ \)};
		\vertex [below=.8of e] (edown){\( \ell^- \)};
		\vertex [right=.7 of b] (b1);
		\vertex [below=1.3 of b1] (b1down) {\(\phantom{Y}\)};
		\feynmandiagram{
			(b) -- (a),
			(b) -- [fermion1](edown),
			(b) -- (e1),
			(eup)-- [fermion1](b),
			(b)-- [opacity=0] (b1down) 
		};
		\end{feynhand}
		\end{feynman}
		\filldraw[fill=black, line width=0.15mm] (1.55,0) circle (3pt);
		\end{tikzpicture}	
		\hspace{1.5cm}
		\begin{tikzpicture}
		\begin{feynman}[large]
		\begin{feynhand}[baseline=(a)]
		\vertex (a);
		\vertex [left=-5cm of a] (xx);
		\vertex [left=0.0cm of a] (a1) {\(X\)};
		\vertex [right=1.5of a] (b);
		\vertex [right=1.4of b] (d);
		\vertex [right=1.2of d] (e);
		\vertex [above=.6of e] (eup){\( \ell^+ \)};
		\vertex [below=.6of e] (edown){\( \ell^- \)};
		\vertex [right=.7 of b] (b1);
		\vertex [below=1.3 of b1] (b1down) {\(Y\)};
		\feynmandiagram{
			(b) -- (a),
			(b) --[boson, edge label'=\(\gamma^\ast \)] (d),
			(d) -- [fermion1](edown),
			(eup)-- [fermion1](d),
			(b)--(b1down)
		};
		\end{feynhand}
		\end{feynman}
		\filldraw[fill=Salmon, line width=0.15mm] (1.3,-0.2) rectangle (1.7,0.2);
		\end{tikzpicture}
		\hspace{1.5cm}
	\begin{tikzpicture}
		\begin{feynman}[large]
		\begin{feynhand}[baseline=(a)]
		\vertex (a);
		\vertex [left=-5cm of a] (xx);
		\vertex [left=0.0cm of a] (a1) {\(X\)};
		\vertex [right=1.5of a, dot] (b) { \(\hphantom{M}\) \ };
		\vertex [right=2.5 of b, dot] (c) { \(\hphantom{M}\) \ };
		\vertex [right=1.4of c] (d);
		\vertex [right=1.2of d] (e);
		\vertex [above=.6of e] (eup){\( \ell^+ \)};
		\vertex [below=.6of e] (edown){\( \ell^- \)};
		\vertex [right=.7 of b] (b1);
		\vertex [below=1.3 of b1] (b1down) {\(Y\)};
		\feynmandiagram{
			(b) -- (a),
			(b)--[quarter left, edge label=\(\pi^+ \)](c), 
			(b)--[quarter right, edge label'=\(\pi^- \)](c), 
			(c) --[boson, edge label'=\(\gamma^\ast \)] (d),
			(d) -- [fermion1](edown),
			(eup)-- [fermion1](d),
			(b)--(b1down)
		};
		\end{feynhand}
		\end{feynman}
		\filldraw[fill=lightgray, line width=0.15mm] (1.8,0) circle [radius=0.3cm];
		\filldraw[fill=white, line width=0.15mm] (4.25,0) circle [radius=0.3cm];
		\end{tikzpicture}
		\hspace{1.5cm}
		\begin{tikzpicture}
		\begin{feynman}[large]
		\begin{feynhand}[baseline=(a)]
		\vertex (a);
		\vertex [left=-5cm of a] (xx);
		\vertex [left=0.0cm of a] (a1) {\(X\)};
		\vertex [right=1.5of a] (b);
		\vertex [right=1.4of b] (d);
		\vertex [right=1.4of d] (d1);
		\vertex [right=1.2of d1] (e);
		\vertex [above=.6of e] (eup){\( \ell^+ \)};
		\vertex [below=.6of e] (edown){\( \ell^- \)};
		\vertex [right=.7 of b] (b1);
		\vertex [below=1.3 of b1] (b1down) {\(Y\)};
		\feynmandiagram{
			(b) -- (a),
			(b) --[double, edge label'=\(\ V \)] (d),
			(d)--[boson, edge label'=\(\gamma^\ast \)] (d1),
			(d1) -- [fermion1](edown),
			(eup)-- [fermion1](d1),
			(b)--(b1down)
		};
		\end{feynhand}
		\end{feynman}
		\filldraw[fill=cyan, line width=0.15mm] (1.3,-0.2) rectangle (1.7,0.2);
		\filldraw[fill=gray, line width=0.15mm] (3,0) circle (3pt);
		\end{tikzpicture}
		} 
		\]
	\caption{Contributions to the $C$- and $CP$-odd decay $X\to Y\ell^+\ell^-$. The first diagram describes a short-range semi-leptonic four-point vertex, the second one includes a long-range hadron--photon coupling, while the last two diagrams account for possible hadronic intermediate states. Among the latter, the pion loop corresponds to an isovector transition while the vector-meson conversion respects the isoscalar part of the virtual photon. The black dot, the red box, the gray  circle, and the blue box
    refer to different $C$- and $CP$-violating vertices, while the white circle is $C$- and $CP$-conserving.}
	\label{fig:feynman_diagrams}
\end{figure}
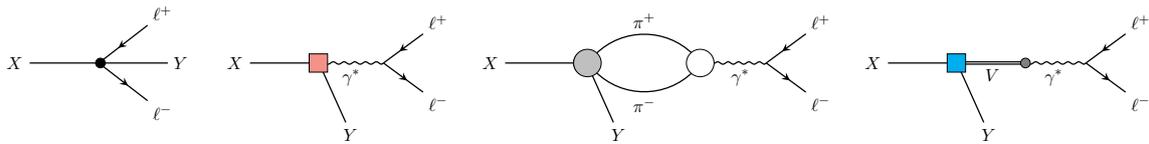
In this section we discuss the phenomenological importance of the three mechanisms driving  $X\to Y\ell^+\ell^-$ and provide the model-independent expressions for them. As an illustration we depict the different contributions in Fig.~\ref{fig:feynman_diagrams}. For simplicity we adapt the notation and conventions introduced for the construction of operators in ToPe$\chi$PT in Ref.~\cite{Akdag:2022sbn} without further details.

\subsection{Kinematics}
\label{sec:kinematics}
Consider the transition amplitude $X(P)\to Y(p)\ell^+(p_{\ell^+})\ell^-(p_{\ell^-})$, with two pseudoscalars $X$ and $Y$ of masses $M_X > M_Y$.
Conventionally, we describe the three-body decay in terms of the Lorentz invariants
\beq
s=(P-p)^2, \qquad t_\ell=(P-p_{\ell^+})^2, \qquad  u_\ell=(P-p_{\ell^-})^2.
\eeq
With the electromagnetic quark current
\beq
 J_\mu=\sum_f Q_f \bar q_f \gamma_\mu q_f\,,
\eeq
where $Q_f$ indicates the electric charge of the respective quarks with flavor $f$ and is conventionally used in units of the proton charge $e$, the singularity-free electromagnetic transition form factor $F_{XY}(s)$
in $X\to Y\gamma^*$ can be decomposed by Poincar\'e invariance and current conservation as~\cite{Bernstein:1965hj, DAmbrosio:1998gur}
\beq\label{eq:formfac}
    \langle Y(p)|J_\mu(0)|X(P)\rangle=-i\left[ s(P+p)_\mu -(P^2-p^2)q_\mu\right]\,F_{XY}(s)\equiv -iQ_\mu\,F_{XY}(s)\,.
\eeq
Here we introduced the photon's momentum $q_\mu=(P-p)_\mu$. Note that $F_{XY}(s)$ thus defined is real at leading order in ToPe$\chi$PT.\footnote{Put differently, $F_{XY}(s)$ is real below the onset of cuts or imaginary parts due to hadronic intermediate states, i.e., for $s\leq 4M_\pi^2$.} Upon contraction with the lepton current the full decay amplitude becomes
\beq\label{eq:MatrixElementLeptonicDecay}
     i\M(X\to Y \ell^+ \ell^-)=e^2 \,(P+p)_\mu F_{XY}(s)\,\Bar{u}_r( p_{\ell^-})\gamma^\mu v_{r'}( p_{\ell^+}),
\eeq
where the term proportional to $q_\mu$ drops out due to current conservation~\cite{Bernstein:1965hj}. In the course of this work, we will see explicitly that the amplitude of each mechanism restores this functional form. Taking the squared absolute value and summing over the lepton spins, one may obtain the doubly differential decay width~\cite{Kubis:2010mp}
\beq
     \frac{\diff\Gamma(X\to Y \ell^+ \ell^-)}{\diff s\, \diff \tau}
     =\frac{\alpha^2}{16\pi M_X^3}  \,\left(\lambda\big(s,M_X^2,M_Y^2\big)-\tau^2\right) \,|F_{XY}(s)|^2,
\eeq
in terms of the electromagnetic fine-structure constant $\alpha=e^2/4\pi$, the K\"all\'en function $\lambda(x,y,z)=x^2+y^2+z^2-2(xy+xz+yz)$, and the Lorentz invariant $\tau=t_\ell-u_\ell$. The $\tau$-integration can be carried out analytically, giving
\beq\label{eq:d_Gamma}
    \frac{\diff\Gamma(X\to Y \ell^+ \ell^-)}{\diff s}
    =\frac{\alpha^2}{8\pi M_X^3}\, \lambda^{3/2}\big(s,M_X^2,M_Y^2\big)\, \sigma_\ell(s)\left(1-\frac{\sigma^2_\ell(s)}{3}\right) \,|F_{XY}(s)|^2,
\eeq
where $\sigma_\ell(s)=\sqrt{1-4m_\ell^2/s}$ and the physical range is restricted to $s\in\left(4m_\ell^2,(M_X-M_Y)^2\right)$. Throughout, we use the masses $m_e=0.51\,\text{MeV}$, $m_\mu=105.66\,\text{MeV}$, $\meta=547.86\,\text{MeV}$, $\metap=957.78\,\text{MeV}$, $\mpip=139.57\,\text{MeV}$, and $\mpii=134.98\,\text{MeV}$~\cite{Workman:2022ynf}. For later use, we also quote the vector-meson masses $M_\omega=782.66\,\text{MeV}$~\cite{Workman:2022ynf} and $M_\rho = 763.7\,\text{MeV}$, where the latter is the real value of  the $\rho(770)$-meson pole  as determined in Ref.~\cite{Garcia-Martin:2011nna}. The errors and additional decimal digits on all of these masses are negligible in our analysis.

\subsection{Direct semi-leptonic contributions to $X\to Y\ell^+\ell^-$}
\label{sec:Direct_Semi-Leptonic}
\begin{sloppypar}
In Ref.~\cite{Akdag:2022sbn} it was shown that the only $C$-odd, $P$-even semi-leptonic four-point vertex to $\eta\to\pi^0\ell^+\ell^-$ at lowest order in the QED fine-structure constant  and soft momenta originates from the dimension-8 LEFT operator
\beq\label{eq:TOPE_LEFT_dim8_semi_leptonic_gluonic}
    \O_{\ell\psi}^{(u)} \equiv \frac{c_{\ell\psi}^{(u)}}{\Lambda^4}\, \bar \ell \gamma^\mu \ell \bar \psi \gamma^\nu T^a \psi G^a_\munu\,,
\eeq
where $c_{\ell\psi}^{(u)}$ denotes flavor-dependent Wilson coefficients.\footnote{In contrast, $CP$-violating quark--lepton operators that contribute to these decays but are $C$-even and $P$-odd already appear at dimension 6~\cite{Sanchez-Puertas:2018tnp,Escribano:2022zgm}.  Note furthermore that this suppression is specific for flavor-diagonal transitions: flavor-changing processes of $C$- and $CP$-odd nature are similarly already generated at dimension 6~\cite{DAmbrosio:1996lam,Shi:2017ffh}.} The choice of the high-energy scale $\Lambda$ depends on the interpretation of the ToPe operators: in the picture of LEFT, $\Lambda$ can be in the order of the electroweak scale, while in the spirit of the Standard Model effective field theory $\Lambda$ is a typical BSM scale. The respective leading ToPe$\chi$PT operators in the large-$N_c$ limit read~\cite{Akdag:2022sbn}
\begin{aligneq}
    \bar X_{\ell\psi}^{(u)}\supset \frac{c^{(u)}_{\ell\psi}}{\Lambda^4} \,\bar g_1^{(u)}\,i\varphi
    &\vev{\lambda_L\partial_\mu \bar{U}^\dagger \bar{U}-\lambda_R \partial_\mu \bar{U} \bar{U}^\dagger}\bar\ell \gamma^\mu \ell\,, \label{eq:Xu}
\end{aligneq}
where we employ the simple single-angle $\eta$-$\eta'$ mixing scheme~\cite{Feldmann:1999uf}, for which the singlet component corresponds to
\beq\label{eq:singlet-field}
     \varphi = \frac{ \sqrt{2} } { 3 \sqrt{3} F_0 }\,  \eta  + \frac{ 4 }{ 3 \sqrt{3} F_0}\, \eta'\,.
\eeq
The meson matrix in the large-$N_c$ limit is then given by
\beq\label{eq:Goldstone_large_Nc}
\bar U=\exp\left(\frac{i\bar \Phi}{F_0}\right)\,, \quad \mathrm{where} \quad  \bar\Phi= 
\begin{pmatrix}
\frac{1}{\sqrt{3}}\eta'+\sqrt{\frac{2}{3}}\eta+\pi^0   & \sqrt{2}\pi^+  &\sqrt{2}K^+   \\
\sqrt{2}\pi^-& \frac{1}{\sqrt{3}}\eta'+\sqrt{\frac{2}{3}}\eta-\pi^0  & \sqrt{2}K^0 \\
\sqrt{2}K^- & \sqrt{2}\bar K^0& \frac{2}{\sqrt{3}}\eta'-\sqrt{\frac{2}{3}}\eta
\end{pmatrix}\,.
\eeq
In both Eqs.~\eqref{eq:singlet-field} and \eqref{eq:Goldstone_large_Nc}, $\eta$ and $\eta'$ refer to the physical fields; see, e.g., Refs.~\cite{Feldmann:1999uf,Kaiser:2000gs,Escribano:2005qq} for detailed discussions on more elaborate mixing schemes.
Furthermore, in relation~\eqref{eq:Xu} we have introduced the spurion matrices $\lambda_{L,R}$ in flavor space, which were defined in Ref.~\cite{Akdag:2022sbn} and acquire the same physical values, namely $\lambda_{L,R}\in\{\text{diag}(1,0,0),\,\text{diag}(0,1,0),\,\text{diag}(0,0,1)\}$ for the quark flavor $\psi=u,d,s$, respectively. 
Besides, $F_0$ denotes the common meson decay constant in the combined chiral and large-$N_c$ limit, $F_0 \lesssim F_\pi \approx 92.3\MeV$.
Summing over $\psi$ and only picking the interactions relevant for our interests, the operator ${\bar X}_{\ell\psi}^{(u)}$ gives rise to the leading-order Lagrangians 
\begin{aligneq}
    \L_{XY\ell^+\ell^-}=\frac{1}{\Lambda^4F_0^2}\,\N_{X\to Y\ell^+\ell^-} \bar \ell \gamma^\mu \ell  X \partial_\mu Y 
\end{aligneq}
with the normalizations
\beq\label{eq:N_XtoYll}
  \N_{\eta\to \pi^0\ell^+\ell^-}\equiv\frac{2\sqrt{2}}{3\sqrt{3}}
    \,\bar g_1^{(u)} \big(c_{\ell u}^{(u)}-c_{\ell d}^{(u)}\big)\,,\qquad \N_{\eta'\to \eta\ell^+\ell^-}\equiv\frac{2\sqrt{2}}{3}\,\bar g_1^{(u)}\big(c_{\ell u}^{(u)}+c_{\ell d}^{(u)}-2c_{\ell s}^{(u)}\big)\,.
\eeq
Both processes are uncorrelated as their normalizations are linearly independent; the flavor combinations reflect the isospin and $SU(3)$ structure of the transitions. Making use of the Dirac equation for the leptons, the corresponding matrix element yields
\beq
    i\M
    =
    e^2(P+p)_\mu F_1(s)\,\bar u_r(p_{\ell^-}) \gamma^\mu v_{r'}(p_{\ell^+})\,, 
\eeq
with
\beq\label{eq:formfac_F1}
    F_1(s)\equiv -\frac{1}{2e^2\Lambda^4F_0^2}\, \N_{X\to Y\ell^+\ell^-} \sim -\frac{1}{\Lambda^4}\frac{2\pi F_0^2}{e^2}
    = - \frac{F_0^2}{2\alpha \Lambda^4}\,.
\eeq
In the last step we applied the NDA assumption $\N_{X\to Y\ell^+\ell^-}\sim 4\pi F_0^4$. Note, however, that the sign of the normalization is not fixed by NDA.
\end{sloppypar}

\subsection{Direct photonic contributions to $X\to Y\gamma^\ast$}
\label{sec:Direct_Photonic}
The leading-order contribution to the effective Lagrangian of $X\to Y\gamma^\ast$ reads~\cite{Akdag:2022sbn}
\beq
\L_{X\to Y\gamma^\ast}=\frac{1}{\Lambda^4F_0^2} \N_{X\to Y\gamma^\ast}\partial_\mu X\partial_\nu Y F^\munu +\O(p^6).
\eeq
We may access the normalization $\N_{X\to Y\gamma^\ast}$ using NDA, by regarding the possible sources on the level of LEFT, cf.\ Ref.~\cite{Akdag:2022sbn}. In this discussion we can directly ignore LEFT sources whose leading-order contributions in ToPe$\chi$PT are proportional to the $\epsilon$-tensor and can thus not generate an even number of pseudoscalars and at the same time preserve parity. The NDA estimate of $\N_{X\to Y\gamma^\ast}$ for the chirality-breaking dimension-7 LEFT quark-quadrilinear~\cite{Khriplovich:1990ef,Conti:1992xn,Engel:1995vv,Ramsey-Musolf:1999cub,Kurylov:2000ub}
\beq\label{eq:LEFT_dim7} 
    \O_{\psi\chi}^{(a)} = \frac{v}{\Lambda^4} c_{\psi\chi}^{(a)}\, \bar\psi \dvec D_\mu \gamma_5\psi\bar\chi\gamma^\mu\gamma_5\chi\,,
\eeq
which is in the focus of Ref.~\cite{Akdag:2022sbn}, yields $evF_0^3/4\pi$, with Higgs vev $v$. For the $C$- and $CP$-odd dimension-8 operators listed in this reference with two quarks and two gluon field strengths, four quarks and one gluon field strength, four quarks and one photon field strength, we have $\N_{X\to Y\gamma^\ast}\sim eF_0^4$, $eF_0^4/(4\pi)$, $F_0^4$, respectively.
It has to be underlined that \emph{each} of these estimates may differ by one order of magnitude, possibly rendering all of these operators to the same numerical size. However, in the scope of these NDA estimations, we assume the normalization of the dimension-7 LEFT operator to dominate the remaining ones. 

Using $\L_{X\to Y\gamma^\ast}$ to evaluate the $C$-odd vertex in the second diagram of Fig.~\ref{fig:feynman_diagrams}, we obtain the matrix element~\cite{Akdag:2022sbn}
\beq\label{eq:amplitude_direct_photonic}
    i\M = e^2(P+p)_\nu F_2(s)\,\Bar{u}_r( p_{\ell^-})\gamma^\nu v_{r'}( p_{\ell^+})\,,
\eeq
with
\beq\label{eq:formfac_F2}
     F_2(s)\equiv\frac{1}{2e\Lambda^4F_0^2}\N_{X \to Y\gamma^\ast}\sim \frac{v F_0}{8\pi\Lambda^4} \,.
\eeq
Again, NDA does not provide any information on the sign of the amplitude.
Comparing Eqs.~\eqref{eq:formfac_F1} and \eqref{eq:formfac_F2}, we note $F_2(s)/F_1(s) \sim \alpha v/(4\pi F_0) \approx 1.5$, hence both contributions are really expected to be of comparable size.

\subsection{Hadronic long-range effects}
\label{sec:hadronic_long_range}
The hadronic long-range contributions to the transition form factor can be constructed with knowledge about ToPe forces in $X\to Y\pi^+\pi^-$~\cite{Akdag:2021efj,Akdag:2023oob}. We consider in the following sections both the isovector and isoscalar part of the photon.

\subsubsection{The isovector contribution}
\label{sec:isovector}
	
\begin{figure}[t!]
		\centering
        \begin{tikzpicture}
\begin{feynman}[large]
\begin{feynhand}[baseline=(a)]
\vertex (a);
\vertex [left=-5cm of a] (xx);
\vertex [left=0.0cm of a] (a1) {\(X\)};
\vertex [right=1.5of a, dot] (b) { \(\hphantom{M}\) \ };
\vertex [right=2.5 of b, dot] (c) { \(\hphantom{M}\) \ };
\vertex [right=1.8of c] (d);
\vertex [right=1.2of d] (e);
\vertex [above=.6of e] (eup){\( \ell^+ \)};
\vertex [below=.6of e] (edown){\( \ell^- \)};
\vertex [right=.7 of b] (b1);
\vertex [below=1.3 of b1] (b1down) {\(Y\)};
\feynmandiagram{
	(a) -- (b),
	(b)--[quarter left, edge label=\(\pi^+ \)](c), 
	(b)--[quarter right, edge label'=\(\pi^- \)](c), 
	(c) --[boson, edge label'=\(\gamma^\ast \)] (d),
 	(d) --[fermion1] (edown),
 	(d) --[fermion1] (eup),
 	(b)--(b1down)
};
\end{feynhand}
\end{feynman}
\filldraw[fill=lightgray, line width=0.15mm] (1.8,0) circle [radius=0.3cm];
\filldraw[fill=white, line width=0.15mm] (4.25,0) circle [radius=0.3cm];
\draw[Red,dashed] (2.5,-0.85) -- (3.5,0.85);
\end{tikzpicture}
		\vspace{.3cm}
		\caption{Discontinuity of the $X\to Y \gamma^*$ transition form factors, representative for the decays $\eta\rightarrow \pi^0 \gamma^*$ and $\eta'\rightarrow \eta \gamma^*$. The white blob denotes the pion vector form factor and the gray one the $C$-violating contributions to the $\eta\to \pi^0\pi^+\pi^-$ and $\eta'\to \eta\pi^+\pi^-$ amplitudes, respectively. The dashed line illustrates the unitarity cut.}
		\label{fig:DiscFormFactor}
	\end{figure}
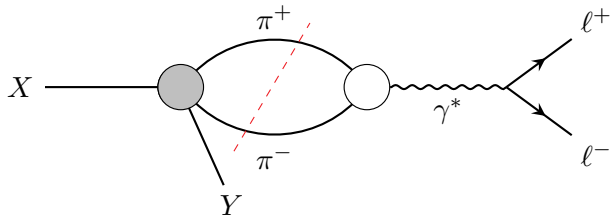

In this section we establish dispersion relations for hadronic contributions of the $C$- and $CP$-odd transition form factor $F_{XY}$ and restrict the calculation to the isovector part of the photon. 
The discontinuity of $X\to Y \gamma^* $, as depicted in Fig.~\ref{fig:DiscFormFactor}, can be calculated by applying a unitarity cut on the dominant intermediate state, i.e., two charged pions, allowing us to access the transition form factor in a non-perturbative fashion.
The first ingredient to the discontinuity in Fig.~\ref{fig:DiscFormFactor} is indicated by the gray blob and describes the $C$- and $CP$-odd contributions to the hadronic $X\to Y\pi^+\pi^-$ decay amplitude defined by
\beq
    \big\langle Y(p)\,\pi^+(p_+)\,\pi^-(p_-)\big|iT\big| X(P)\big\rangle
    =(2\pi)^4\,\delta^{(4)}( P-p-p_+-p_-)\,i\M^{XY}(s,t,u)\,.
\eeq
These amplitudes will be discussed in detail for the different cases in Sect.~\ref{sec:Hadronic_long_range_isovector}.
The remaining contribution is the 
pion vector form factor defined via the current\footnote{In the isospin limit, which we will employ for the pion form factor in the following, only the isovector contribution of the current contributes, i.e., $J^{(1)}_\mu=\frac{1}{2}\big(\bar u \gamma_\mu u-\bar d \gamma_\mu d\big)$.}
\beq
\langle \pi^+(p_+)\pi^-(p_-)|J_\mu(0)|0\rangle = (p_+-p_-)_\mu \Fpivec(s)\,.
\eeq 
Note that this equation and Eq.~\eqref{eq:formfac} differ, beside the respective momentum configuration, by an explicit imaginary unit as demanded by their different behavior under time reversal.
With only elastic rescattering taken into account, the pion vector form factor obeys the discontinuity relation
\beq
    \disc F^V_\pi(s) = 2i\,F^V_\pi(s)\sin\delta_1(s)e^{-i\delta_1(s)} \theta(s-4M_\pi^2)\,,
\eeq
where $\delta_1(s)$ denotes the $P$-wave $\pi\pi$ phase shift with two-body isospin $I_{\pi\pi}=1$. The most general solution to this equation is given in terms of the Omn\`es function~\cite{Omnes:1958hv}  
\beq
    F^V_\pi(s)= P_n(s)\,\Omega_1(s)=P_n(s)\,\mathrm{exp}\Bigg( \frac{s}{\pi} \int\limits_{4M_\pi^2}^\infty \  \frac{\delta_1(x)}{x(x-s)}  \ \mathrm{d}x  \Bigg),
\eeq
with a real-valued subtraction polynomial $P_n$ of order $n$. The index of the Omn\`es function indicates the isospin $I_{\pi\pi}$ of the dipion state. The pion vector form factor is expected to behave as $F_\pi^V(s) \asymp 1/s$ for large energies~\cite{Chernyak:1977as,Chernyak:1980dj,Efremov:1978rn,Efremov:1979qk,Farrar:1979aw,Lepage:1979zb,Lepage:1980fj,Leutwyler:2002hm} (up to logarithmic corrections), and to be free of zeros~\cite{Leutwyler:2002hm,Ananthanarayan:2011xt}. Thus, $P_n$ is a constant and can be set to $1$ due to gauge invariance, such that $\Fpivec(s)=\Omega_1(s)$. For consistency we waive the incorporation of inelastic effects, which we do not consider in $X\to Y \pi^+\pi^-$ either. In the region of the $\rho(770)$ resonance dominating $F^V_\pi(s)$, these are known to affect the form factor by no more than 6\%, depending on the phase shift used as input~\cite{Hanhart:2016pcd}.  Given other sources of uncertainty in the present study, we consider this error negligible.

When we cut the dipion intermediate state in Fig.~\ref{fig:DiscFormFactor}, the discontinuity of the isovector contribution $F^{(1)}_{XY}$ to the transition form factor $F_{XY}$ becomes 
\beq
    Q_\mu\text{disc}F^{(1)}_{XY}(s)=\int\frac{\diff^4k}{(2\pi)^2}\,\delta(k^2-\mpi^2)\,\delta\left( (q-k^2)-\mpi^2 \right) \,\M^{XY}(s,t,u)\,(q-2k)_\mu \Fpiveccon(s)\,,
\eeq
where $t=(P-p_+)^2$ and $u=(P-p_-)^2$.  We find
\beq\label{eq:disc}
       \text{disc}F^{(1)}_{XY}(s)=-\frac{1}{24\pi}\sigma^3_\pi(s)\Fpiveccon(s)\,f_{XY}(s)\,\theta(s-4\mpi^2)\,,
\eeq
where $\sigma_\pi(s)=\sqrt{1-4\mpi^2/s}$.
In this discontinuity relation the quantity $f_{XY}$ denotes the $P$-wave projection of the hadronic decay amplitude given by 
\beq\label{eq:definition_partial_wave}
    f_{XY}(s)\equiv \frac{3}{2\kappa(s)}\int_{-1}^1 \diff z\, z\, \M^{XY}(s,t,u),
\eeq
with 
\beq \label{eq:zs}
z=\frac{t-u}{\kappa(s)}\eqand \kappa(s)=\sigma_\pi(s)\lambda^{1/2}\big(s, M_X^2,M_Y^2\big)\,. 
\eeq
Adapting the high-energy behavior of $f_{XY}$ and $\delta_1$ from Refs.~\cite{Akdag:2021efj,Akdag:2023oob}, an unsubtracted dispersion relation is sufficient to ensure convergence of the remaining integral over the discontinuity, such that the form factor can be evaluated with
\beq\label{eq:isovector_formfactor}
    F^{(1)}_{XY}(s)=\frac{i}{48\pi^2}\int_{4\mpi^2}^{\infty}\diff x\, \sigma^3_\pi(x)\Fpiveccon(x)\,\frac{f_{XY}(x)}{x-s}.
\eeq

\boldmath
\subsubsection{The isoscalar contribution}
\label{sec:isoscalar}
\begin{figure}
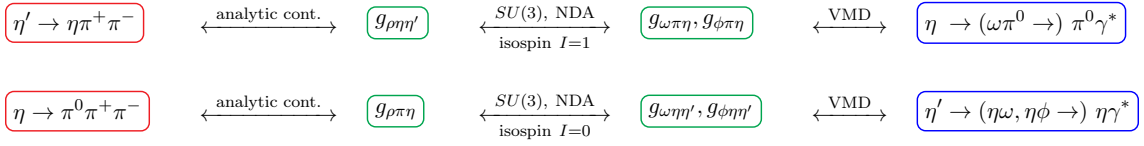

    \centering
    \begin{equation}
    \nonumber
    \resizebox{\textwidth}{!}{
    $
    \begin{aligned}
    &\RedBox{\eta'\to\eta\pi^+\pi^-} \quad
    &&\xleftrightarrow{\ \text{analytic cont.} \ }  \quad
    \GreenBox{g_{\rho\eta\eta'}}  \quad
    &&\xleftrightarrow[\text{isospin}\ I=1]{\ SU(3),\  \text{NDA} \ }  \quad
    \GreenBox{g_{\omega\pi\eta}, g_{\phi\pi\eta}} \quad
    &&\xleftrightarrow{\ \text{VMD} \ }  \quad
    \BlueBox{\eta\phantom{'}\to (\omega\pi^0\to)\ \pi^0\gamma^\ast}\\[0.5cm]
    &\RedBox{\eta\to\pi^0\pi^+\pi^-} \quad
    &&\xleftrightarrow{\ \text{analytic cont.} \ }  \quad
    \GreenBox{g_{\rho\pi\eta}}  \quad
    &&\xleftrightarrow[\text{isospin}\ I=0]{\ SU(3),\  \text{NDA}\ }  \quad
    \GreenBox{g_{\omega\eta\eta'}, g_{\phi\eta\eta'}} \quad
    &&\xleftrightarrow{\ \text{VMD} \ }  \quad
    \BlueBox{\eta'\to (\eta\omega, \eta\phi\to)\ \eta\gamma^\ast}
    \end{aligned}
    $
   }
   \end{equation}
    \caption{Schematic sketch of the strategy to extract the isoscalar contribution of the $\eta\to\pi^0\gamma^\ast$ ($\eta'\to\eta\gamma^\ast$) transition form factor (blue) from the $C$- and $CP$-odd $\eta'\to\eta\pi^+\pi^-$ ($\eta\to\pi^0\pi^+\pi^-$) amplitude (red) using vector-meson couplings (green).}
    \label{fig:couplings}
\end{figure}

\unboldmath
In order to estimate the isoscalar contribution, we apply a VMD pole approximation and consider a vector-meson conversion of $\gamma^*$ to $v_\mu$, with $v\in \{\omega, \phi\}$, cf.\ the very right diagram in Fig.~\ref{fig:feynman_diagrams}. While this strategy is not as model-independent and sophisticated as the dispersive analysis of the isovector part of $\gamma^*$, it serves as a good approximation to at least estimate the relative size of this contribution, 
not least due to the narrowness of the $\omega$ and $\phi$ resonances dominating isoscalar vector spectral functions at low energies. Furthermore, this ansatz even correlates the decay $\eta\to \pi^0\gamma^\ast$ to $\eta'\to \eta\pi^+\pi^-$ and $\eta'\to \eta\gamma^\ast$ to $\eta\to \pi^0\pi^+\pi^-$ by following the strategy sketched in Fig.~\ref{fig:couplings} to relate the decays of same total isospin.

The combination of vector mesons with $\chi$PT was extensively worked out for instance in Refs.~\cite{Lee:1972epa, Kaymakcalan:1984bz, Meissner:1987ge, Jain:1987sz, Klingl:1996by, Mai:2022eur} and references therein.
The number of free parameters can be reduced most efficiently, cf.\ Ref.~\cite{Meissner:1987ge}, by coupling uncharged vector mesons to uncharged pseudoscalars via the field-strength tensor $V_{L,R}^\munu$. The latter is the analog to the photonic one with the same discrete symmetries and transformations under $SU(3)_L\times SU(3)_R$. If we only consider the relevant degrees of freedom, i.e., treating $\bar U$ and $V_{L,R}^\mu$ as diagonal matrices, we can effectively write
\beq\label{eq:field_strength}
    V_{L,R}^\munu=\partial^\mu V^\nu_{L,R}-\partial^\nu V^\mu_{L,R}\,.
\eeq
The physical value of this chiral building block can be evaluated with $V_{L}^\mu=V_{R}^\mu=\diag(\rho+\omega, -\rho+\omega, \sqrt{2}\phi)^{\mu}+\ldots$, where the ellipsis indicates terms without vector mesons. 
At the mesonic level, we can deduce the desired interaction from the leading-order $XY\gamma^{*}$ operator, cf.\ Ref.~\cite{Barrett:1965ia}, and hence write
\beq\label{eq:generic_vector_meson_Lagrangian}
    \L_{vYX}=g_{vYX}\partial_\mu X\partial_\nu Yv^\munu\,,
\eeq
with $v_\munu\equiv\partial_\mu v_\nu-\partial_\nu v_\mu$. The ToPe$\chi$PT operators that can generate this mesonic interaction at leading order in the large-$N_c$ power counting, i.e., $\O(p^4,\delta^2)$ (see Ref.~\cite{Akdag:2022sbn} for further details), and originate from the LEFT operator in Eq.~\eqref{eq:LEFT_dim7} are
\begin{aligneq}\label{eq:vector_meson_example_ToPeChPT}
    \bar{X}_{\psi\chi}^{(a)}\supset \frac{v}{\Lambda^4}c_{\psi\chi}^{(a)} 
    \Big(
    &\bar{g}_{V_1}^{(a)}\vev{\lambda_L\partial_\nu \bar U^\dagger \bar U-\lambda_R\partial_\nu \bar U\bar U^\dagger}\vev{\big(\lambda V_L^\munu \partial_\mu \bar U^\dagger-\lambda^\dagger V_R^\munu \partial_\mu \bar U\big)-\hc}\\[0.1cm]
    &+\bar{g}_{V_2}^{(a)}\vev{\lambda \partial_\mu \bar U^\dagger-\lambda^\dagger  \partial_\mu \bar U}\vev{\big(\lambda_L V_L^\munu\partial_\nu \bar U^\dagger \bar U-\lambda_R V_R^\munu\partial_\nu \bar U\bar U^\dagger\big)-\hc}\\[0.1cm]
    &+\bar{g}_{V_3}^{(a)} \partial_\nu \varphi\vev{\big(\lambda\lambda_L V_L^\munu\partial_\mu \bar U^\dagger -\lambda^\dagger\lambda_R V_R^\munu\partial_\mu \bar U\big)-\hc}
    \Big)
   \,.
\end{aligneq}
Here, we only list the operators leading to distinct, non-vanishing, expressions after we set $\lambda^{(\dagger)}$, $\lambda_{L,R}$, and $V^\munu_{L,R}$ to their physical values and use the fact that in our application all appearing matrices are diagonal and therefore commute. Evaluating the flavor traces of the operator in the first line and labeling the vector-meson couplings with a corresponding superscript $(V_1)$, we end up with
\begin{aligneq}\label{eq:uncorrelated_couplings}
        &g^{(V_1)}_{\rho\pi\eta}&&=\frac{16v}{\Lambda^4F_0^2}\sqrt{\frac{2}{3}}\big(c_{ud}^{(a)}+c_{du}^{(a)}\big)&&\bar{g}_{V_1}^{(a)}-\frac{1}{\sqrt{3}}g^{(V_1)}_{\omega\eta\eta'} \,,
        \qquad
        &&
        g^{(V_1)}_{\rho\eta\eta'}&&=\hphantom{-}\frac{8\sqrt{2}v}{\Lambda^4F_0^2}\big(c_{us}^{(a)}-c_{ds}^{(a)}\big)&&\bar{g}_{V_1}^{(a)} \,,
        \\
        &g^{(V_1)}_{\omega\pi\eta}&&=\frac{16v}{\Lambda^4F_0^2}\sqrt{\frac{2}{3}}\big(c_{ud}^{(a)}-c_{du}^{(a)}\big)&&\bar{g}_{V_1}^{(a)} -\frac{1}{\sqrt{3}} g^{(V_1)}_{\rho\eta\eta'}\,,
        &&
        g^{(V_1)}_{\omega\eta\eta'}&&=\hphantom{-}\frac{8\sqrt{2}v}{\Lambda^4F_0^2}\big(c_{us}^{(a)}+c_{ds}^{(a)}\big)&&\bar{g}_{V_1}^{(a)} \,,
        \\
        &g^{(V_1)}_{\phi\pi\eta}&&=\frac{16v}{\sqrt{3}\Lambda^4F_0^2}\big(c_{su}^{(a)}-c_{sd}^{(a)}\big)&&\bar{g}_{V_1}^{(a)}\,, 
        &&g^{(V_1)}_{\phi\eta\eta'}&&=-\frac{16v}{\Lambda^4F_0^2}\big(c_{su}^{(a)}+c_{sd}^{(a)}\big)&&\bar{g}_{V_1}^{(a)} \,.
\end{aligneq}
For the second operator in Eq.~\eqref{eq:vector_meson_example_ToPeChPT} we observe that the resulting vector meson couplings $g^{(V_2)}_{vYX}$ equal the $g^{(V_1)}_{vYX}$ from Eq.~\eqref{eq:uncorrelated_couplings} if $c_{\psi\chi}^{(a)}\bar{g}_{V_2}^{(a)}  =-c_{\chi\psi}^{(a)} \bar{g}_{V_1}^{(a)} $. The third operator in $\bar X_{\psi\chi}^{(a)}$ yields 
\begin{aligneq}\label{eq:correlated_couplings}
        &g^{(V_3)}_{\rho\pi\eta}&&=\frac{1}{\sqrt{3}}g^{(V_3)}_{\omega\eta\eta'} \,,
        \qquad
        &&
        g^{(V_3)}_{\rho\eta\eta'}&&=-\frac{4\sqrt{2}v}{3\Lambda^4F_0^2}\big(c_{uu}^{(a)}-c_{dd}^{(a)}\big)&&\bar{g}_{V_3}^{(a)} \,,
        \\
        &g^{(V_3)}_{\omega\pi\eta}&&=\frac{1}{\sqrt{3}} g^{(V_3)}_{\rho\eta\eta'}\,,
        &&
        g^{(V_3)}_{\omega\eta\eta'}&&=-\frac{4\sqrt{2}v}{3\Lambda^4F_0^2}\big(c_{uu}^{(a)}+c_{dd}^{(a)}\big)&&\bar{g}_{V_3}^{(a)} \,,
        \\
        &g^{(V_3)}_{\phi\pi\eta}&&=0\,, 
        &&g^{(V_3)}_{\phi\eta\eta'}&&=\phantom{-}\frac{16v}{3\Lambda^4F_0^2}c_{ss}^{(a)}\bar{g}_{V_3}^{(a)} \,.&&
\end{aligneq}
Both Eqs.~\eqref{eq:uncorrelated_couplings} and \eqref{eq:correlated_couplings} suggest that there is a correlation between $g_{\rho\pi\eta}$ and $g_{\omega\eta\eta'}$ as well as between $g_{\rho\pi\eta}$ and $g_{\omega\eta\eta'}$, but none of $g_{\phi\pi\eta}$ or $g_{\phi\eta\eta'}$ with the $\rho$ couplings. However, this observation does not necessarily hold for higher orders in ToPe$\chi$PT or for operators derived from other LEFT sources.  
We continue with the couplings in Eq.~\eqref{eq:correlated_couplings} as our central estimates and make use of the flavor relations implied therein.  

Next, we consider the Lagrangian 
\beq \label{eq:OmegaInteractions}
    \L_{v\gamma}=-\frac{eM^2_v}{g_{v\gamma}} A_\nu v^\nu\,
\eeq 
for the vector-meson conversion with known coupling constants $g_{v\gamma}$.
As Eq.~\eqref{eq:OmegaInteractions} employs the photon field instead of the field strength tensor, it is not manifestly gauge invariant. In the end, this is a necessity to implement \emph{strict} VMD for the isoscalar part of the form factor, avoiding an additional direct photon coupling.
We can now evaluate the isoscalar contribution illustrated on the very right in Fig.~\ref{fig:feynman_diagrams}, which, in agreement with  Eq.~\eqref{eq:MatrixElementLeptonicDecay}, gives rise to the matrix element
\beq
    i\M^{(0)}_{XY}=
    e^2\,  (P+p)_\mu\, F^{(0)}_{XY}(s)\,
    \Bar{u}_r( p_{\ell^-})\gamma^\mu v_{r'}( p_{\ell^+})\,.
\eeq
The corresponding isoscalar form factor, which is consistent with the high-energy behavior of the isovector part in Eq.~\eqref{eq:isovector_formfactor}, finally reads 
\beq\label{eq:FormFacVMD}
    F_{XY}^{(0)}(s)\equiv
    \frac{\tilde g}{2g_{v\gamma}}\frac{M^2_v}{M^2_v-s}\,,
\eeq
where $\tilde g$ equals $g_{v\pi\eta}$ for $X=\eta$, $Y=\pi^0$ and $g_{v\eta\eta'}$ for $X=\eta'$, $Y=\eta$.

\subsection{Discussion}
\label{sec:discussion_phenomenology}

With the results worked out in the previous sections we can evaluate the full contribution of the $X\to Y\gamma^\ast$ transition form factors by
\beq\label{eq:form_factor_full_contribution}
    F_{XY}(s)= F_1(s)+F_2(s)+F_{XY}^{(1)}(s)+F_{XY}^{(0)}(s)\,,
\eeq
where each summand corresponds to one diagram in Fig.~\ref{fig:feynman_diagrams}.
Note that there is no way to distinguish between the four contributions in a sole measurement of the $X\to Y\ell^+\ell^-$ branching ratio. 

Regarding $F_1$ and $F_2$, we observe that their NDA estimates in Eqs.~\eqref{eq:formfac_F1} and \eqref{eq:formfac_F2} yield roughly the same result, even without accounting for the uncertainty of NDA. Hence, there is no clear hierarchy between direct semi-leptonic contributions and $C$- and $CP$-violating photon--hadron couplings contributing to $X\to Y\ell^+\ell^-$. 
In future analyses, the sum $F_1+F_2$ (which does not depend on $s$ at leading order in ToPe$\chi$PT) may be replaced by a single constant parameter in a regression to hypothetical measurements of respective singly- or doubly-differential momentum distributions.

We remark in passing that \emph{all} transition form factor contributions could be expected to undergo further hadronic corrections due to ``initial-state interactions'' of $\eta\pi$ and $\eta'\eta$ $P$-wave type, respectively.  However, all corresponding phase shifts are expected to be tiny and the resulting effects to be hence utterly negligible: the $\eta\pi$ $P$-wave is strongly suppressed in the chiral expansion at low energies relative to $\eta\pi$ $S$-wave or $\pi\pi$ rescattering~\cite{Bernard:1991xb, Kubis:2009sb}, and resonances with quark-model-exotic quantum numbers $J^{PC}=1^{-+}$ due to their $C$-odd nature will have a rather large mass~\cite{JPAC:2018zyd}.  We therefore do not consider any such corrections in this article.

Given the currently accessible experimental data and the missing information on the normalizations of $F_1$ and $F_2$, we henceforth focus on the contributions of $F_{XY}^{(1)}$ and $F_{XY}^{(0)}$. On the one hand, we are in a position to predict the latter with the input discussed in Sect.~\ref{sec:hadronic_long_range}. On the other hand, they provide new conceptual insights by directly relating ToPe forces in $X\to Y\pi^+\pi^-$ with $X\to Y\gamma^\ast$. Moreover, for simplicity we assume no significant cancellations among the individual contributions in Eq.~\eqref{eq:form_factor_full_contribution} throughout this manuscript.  
We leave the study of such a more complicated interplay between different mechanisms for future analyses, when more rigorous experimental bounds may be exploited for correlated constraints between them.

\boldmath
\section{Hadronic long-range effects: the isovector contribution}
\label{sec:Hadronic_long_range_isovector}
\unboldmath
\begin{sloppypar}
In this section we investigate the isovector contribution to the transition form factor $X\to Y\gamma^\ast$ based on the dispersive representations derived in Sect.~\ref{sec:isovector}. Thus, we focus on $C$- and $CP$-odd contributions of the lowest-lying hadronic intermediate state, i.e., on the decay chain $X\to Y \pi^+\pi^- \to Y\gamma^\ast$.
\end{sloppypar}

\boldmath
\subsection{The dispersive $C$- and $CP$-odd $X\to Y\pi^+\pi^-$ partial-wave amplitude}
\unboldmath
\label{sec:KT_Eta}

The formalism, results, and most of the notation are adopted from Refs.~\cite{Akdag:2021efj,Akdag:2023oob}.
The latter uses a dispersive framework known as Khuri--Treiman equations~\cite{Khuri:1960zz} to access the three-body amplitude $X\to Y\pi^+\pi^-$ including its $C$- and $CP$-odd contributions. In this approach, a coupled set of integral equations is set up for the two-body scattering process and analytically continued to the physical realm of the three-body decay.

\boldmath
\subsubsection{$\eta\to\pi^0\pi^+\pi^-$}
\label{sec:Eta3Pi_partial_wave}
\unboldmath

The $C$- and $CP$-odd contributions to $\eta\to\pi^0\pi^+\pi^-$ read
\beq\label{eq:FullAmpEta3Pi}
    \M^{\eta\pi}(s,t,u)= \M^{\eta\pi}_0(s,t,u) + \M^{\eta\pi}_2(s,t,u), 
\eeq
where the lower index denotes the \emph{total} isospin of the three-body final state.
Neglecting $D$- and higher partial waves, we can decompose these amplitudes in the sense of a reconstruction theorem~\cite{Stern:1993rg,Ananthanarayan:2000cp,Zdrahal:2008bd} into single-variable functions $\G_{I_{\pi\pi}}(s)$, $\H_{I_{\pi\pi}}(s)$ with fixed \emph{two-body} isospin $I_{\pi\pi}$ and relative angular momentum $\ell$ of the $\pi^+\pi^-$ state:
\beq
\begin{split}
    \M^{\eta\pi}_0(s,t,u)&=(t-u)\G_1(s) + (u-s)\G_1(t)+(s-t)\G_1(u),
    \\
	\M^{\eta\pi}_2(s,t,u)&=2(u-t)\H_1(s) + (u-s)\H_1(t)+(s-t)\H_1(u)-\H_2(t)+\H_2(u) \vphantom{\frac{2}{3}}.
\end{split} \label{eq:RT-eta3pi}
\eeq
Due to Bose symmetry the $I_{\pi\pi}=1$ single-variable functions have $\ell=1$ while the ones with $I_{\pi\pi}=2$ have $\ell=0$. Unitarity demands the single-variable functions to obey
\beq\label{eq:DiscKT}
    \disc \A_{I_{\pi\pi}}(s) = 2i\,\sin\delta_{I_{\pi\pi}}(s)e^{-i\delta_{I_{\pi\pi}}(s)} \left(\A_{I_{\pi\pi}}(s)+\hat \A_{I_{\pi\pi}}(s)\right)\theta(s-4M_\pi^2),
\eeq
with $\A_{I_{\pi\pi}}\in\{\G_{I_{\pi\pi}},\H_{I_{\pi\pi}}\}$ and the $\pi\pi$ scattering phase shift $\delta_{I_{\pi\pi}}(s)$. The inhomogeneity $\hat\A_{I_{\pi\pi}}(s)$ contains left-hand-cut contributions induced by crossed-channel rescattering effects. In terms of the angular average
\beq\label{eq:AngAver}
    \langle z^n f\rangle :=\frac{1}{2}\int_{-1}^{1}\diff z\,z^n f\!\left(\textstyle\frac{3r-s+z\kappa(s)}{2}\right),
\eeq
with $3r\equiv s+t+u$, the $\hat\A_{I_{\pi\pi}}(s)$ explicitly read
\beq\label{eq:InhomEta3PiCvio}
\begin{split}
	\hat\G_1(s)&=-\frac{3}{\kappa(s)} \left( 3(s-r) \langle z\G_1 \rangle +\kappa(s) \langle z^2 \G_1 \rangle \right),
	\\
	\hat\H_1(s)&=\frac{3}{2\kappa(s)} \left( 3(s-r) \langle z\H_1 \rangle +\kappa(s) \langle z^2 \H_1 \rangle +2\langle z\H_2 \rangle \right),
	\\
	\hat\H_2(s)&=\frac{1}{2} \left( 9(s-r) \langle \H_1 \rangle +3\kappa(s) \langle z \H_1 \rangle -2\langle \H_2 \rangle \right).
\end{split}
\eeq
For the $\A_{I_{\pi\pi}}(s)$ we employ dispersion relations with a minimal number of subtractions to ensure convergence. Assuming that in the limit of infinite $s$ the phase shifts scale like $\delta_1(s)\to \pi$, $\delta_2(s)\to 0$ and the single-variable functions as $\A_1(s)=\mathcal{O}(s^{-1})$,  $\A_2(s)=\mathcal{O}(s^0)$, we obtain 
\beq\label{eq:SubSchemeEta3PiCvio}
    \begin{split}
			\G_1(s)&= \Omega_1(s)\Biggl(\varepsilon+\frac{s}{\pi}\int\limits_{4M_\pi^2}^\infty 
			\frac{\diff x}{x}\, \frac{\sin\delta_1(x)\hat{\G}_1(x)}{|\Omega_1(x)|\, (x-s)}\Biggr),\\
			\H_1(s)&= \Omega_1(s)\Biggl(\vartheta+\frac{s}{\pi}\int\limits_{4M_\pi^2}^\infty 
			\frac{\diff x}{x}\, \frac{\sin\delta_1(x)\hat{\H}_1(x)}{|\Omega_1(x)|\, (x-s)}\Biggr),\\
			\H_2(s)&= \Omega_2(s)\frac{s}{\pi}\int\limits_{4M_\pi^2}^\infty 
			\frac{\diff x}{x}\, \frac{\sin\delta_2(x)\hat{\H}_2(x)}{|\Omega_2(x)|\, (x-s)}\,,
	\end{split}
\eeq
where here and in the following $\mpi\equiv\mpip$.
The $C$-conserving SM amplitude for $\eta\to\pi^+\pi^-\pi^0$ is similarly described in terms of Khuri--Treiman amplitudes; these have been discussed extensively in the literature, see Ref.~\cite{Colangelo:2018jxw} and references therein.
The subtraction constants obtained by a fit to the Dalitz-plot distributions\footnote{The latest BESIII data for $\eta\to\pi^+\pi^-\pi^0$~\cite{BESIII:2023edk} is not included in Refs.~\cite{Akdag:2021efj,Akdag:2023oob} yet and has also not been added to our present analysis, as the statistical accuracy does not supersede that of Ref.~\cite{Anastasi:2016cdz}.} of $X\to Y\pi^+\pi^-$~\cite{Anastasi:2016cdz} yield~\cite{Akdag:2021efj,Akdag:2023oob}
\beq\label{eq:subcons_eta}
    \varepsilon= i\, 0.014(22)\,\mpi^{-2}\,,\qquad \vartheta= i\, 0.068(34)\times 10^{{-3}}\,\mpi^{-2}\,.
\eeq 
These subtraction constants give rise to the real-valued isoscalar and isotensor couplings $g_0=-2.8(4.5)\,\text{GeV}^{-6}$ and $g_2=-9.3(4.6)\cdot 10^{-3}\,\text{GeV}^{-2}$~\cite{Akdag:2021efj,Akdag:2023oob}, using 
\beq\label{eq:subcons_couplings_eta}
   \varepsilon= -0.25i\,\text{GeV}^4\,g_0\,,\qquad\quad
   \vartheta  = -0.38i\,g_2\,.
\eeq
 With the $\A_{I_{\pi\pi}}(s)$ defined above, the $P$-wave amplitude necessary to evaluate the $\eta\to\pi^0\gamma^*$ transition form factor is by definition, cf.\ Eq.~\eqref{eq:definition_partial_wave}, given as
\beq
    f\etapin(s)=\G_1(s)+\hat\G_1(s)+\H_1(s)+\hat\H_1(s)\,,
\eeq
whose dependence on $\H_2$ and $\hat\H_2$ is rather subtle and enters the definition of $\hat\H_1$ in Eq.~\eqref{eq:InhomEta3PiCvio}.

The transition form factor is fully determined by knowledge about the partial-wave amplitude $f_{XY}(s)$ and the pion vector form factor $\Fpivec(s)$. These quantities are in turn fixed by the subtraction constants $\varepsilon$, $\vartheta$,
the $S$-wave $\pi\pi$ scattering phase shift  $\delta_2$ with isospin $2$, and 
the $P$-wave $\pi\pi$ scattering phase shift 
$\delta_1$~\cite{Caprini:2011ky},
respectively. The latter has to be used consistently in $f_{XY}(s)$ and $\Fpivec(s)$, i.e., we use the same continuation to asymptotic $s$ and omit the incorporation of inelasticities, which is beyond the scope of this work.

\boldmath
\subsubsection{$\eta'\to\eta\pi^+\pi^-$}
\label{sec:EtapEtaPiPi_partial_wave}
\unboldmath

In the $C$-and $CP$-violating contribution to $\eta'\to\eta\pi^+\pi^-$ the three-body final state carries total three-body isospin $1$. The respective amplitude can be decomposed as
\beq
    \M^{\eta'\eta}(s,t,u) = (t-u)\G\pipi(s) + \G\etapi(t) - \G\etapi(u);
\eeq
see Ref.~\cite{Isken:2017dkw} for the corresponding SM amplitude.
The indices labeling the single-variable functions indicate which two particles contribute to the intermediate state of the scattering process. While the $\pi\pi$ intermediate state has the quantum numbers $I_{\pi\pi}=1,\,\ell=1$, $\eta\pi$ has $I_{\eta\pi}=1,\,\ell=0$. Both $\G\pipi(s)$ and $\G\etapi(s)$ fulfill the discontinuity equation as quoted in Eq.~\eqref{eq:DiscKT}. The inhomogeneities in this case are
\beq\label{eq:InhomEtapEtaCvio}
\begin{split}
    \hat\G\pipi(s)&=\frac{6}{\kappa\pipi}\, \langle z_s\G\etapi \rangle, \\
    \hat\G\etapi(t)&=-\langle \G\etapi \rangle^+ -\frac{3}{2}\left( r-t+\frac{\Delta}{3t} \right)\langle  \G\pipi \rangle^- + \frac{1}{2}\kappa\etapi\langle z_t\G\pipi \rangle^-,
\end{split}
\eeq
where the cosine of the scattering angles in the $s$-channel is still given by the general expression Eq.~\eqref{eq:zs}, $z_s \equiv z$, 
while the one in the $t$-channel reads
\beq
    z_t=\frac{t\,(u-s)-\Delta}{t\,\kappaeta(t)}\eqwith \kappaeta(t)=\frac{\lambda^{1/2}(t,\metap^2,\mpi^2)\,\lambda^{1/2}(t,\meta^2,\mpi^2)}{t}\,.
\eeq
In these equations we used the notation $\Delta\equiv ( \metap^2-\mpi^2 )( \meta^2-\mpi^2 )$.
We additionally introduced two new types of angular averages, namely
\begin{align}
 \langle z^n f\rangle^{\pm} &:=\frac{1}{2}\int_{-1}^{1}\diff z\,z^n f\!\left(\textstyle\frac{3r-t+z\kappa\etapi(t)\pm\Delta/t}{2}\right)\,.
\end{align}
Assuming the asymptotics $\delta\etapi(t)\to \pi$ and $ \G\pipi(s)=\mathcal{O}(1/s)$, $\G\etapi(t)=\mathcal{O}(t^0)$, the single-variable functions can be evaluated by 
\beq
\begin{split}
			\G\pipi(s)&= \Omega\pipi(s)\left(\varrho +\frac{s}{\pi}\int\limits_{\sth}^\infty \frac{\diff x}{x} \frac{\sin\delta\pipi(x)\hat{\G}\pipi(x)}{|\Omega\pipi(x)|\, (x-s)}\right),\\
			\G\etapi(t)&= \Omega\etapi(t)\left(\zeta \, t+\frac{t^2}{\pi}\int\limits_{\tth}^\infty \frac{\diff x}{x^2} \frac{\sin\delta\etapi(x)\hat{\G}\etapi(x)}{|\Omega\etapi(x)|\,(x-t)}\right).
\end{split}
\eeq
Here, the $\eta\pi$ $S$-wave phase shift from Refs.~\cite{Albaladejo:2015aca,Lu:2020qeo} has been employed.
For the subtraction constants, the values
\beq\label{eq:subcons_eta'}
    \varrho= -i\,0.04(12)\,\mpi^{-2}\,,\qquad \zeta= i\,0.05(12)\,\mpi^{-2}\,,
\eeq
have been obtained by a regression to the Dalitz-plot distribution of $\eta'\to \eta\pi^+\pi^-$~\cite{BESIII:2017djm}. In terms of the real-valued isovector coupling $g_1=0.7(1.0)\,\text{GeV}^{-2}$ and its leading correction $\delta g_1=-5.5(7.3)\,\text{GeV}^{-2}$~\cite{Akdag:2023oob}, the subtraction constants read
\beq\label{eq:subcons_couplings_eta'}
   \varrho= -3.5\cdot 10^{-3}\,ig_1 \left(1-166.5\,\text{GeV}^{2}\,\delta g_1\right)\,,\qquad\quad
   \zeta=  0.76\,ig_1 \left(1-0.65\,\text{GeV}^{2} \, \delta g_1\right)\,.
\eeq
Finally, the $P$-wave entering the $\eta'\to\eta\gamma^*$ transition form factor is given by
\beq
    f\etapeta(s)=\G\pipi(s)+\hat\G\pipi(s).
\eeq
The dependence of $f\etapeta$ on the $S$-wave amplitude $\G\etapi$ is encoded in the angular averages in Eq.~\eqref{eq:InhomEtapEtaCvio}.

\boldmath
\subsection{Computation of the isovector form factor $X \to Y \gamma^\ast$}
\unboldmath
\label{sec:Formfac_computation}
When computing the transition form factors $F^{(1)}_{XY}$, it is advantageous to exploit the linearity of the three-body decay amplitudes $\M^{XY}$ in the subtraction constants. As mentioned in Refs.~\cite{Akdag:2021efj,Akdag:2023oob}, the solutions of the Khuri--Treiman amplitudes can be represented by so-called basis solutions, which are independent of the subtraction constants and can be fixed once and for all before even carrying out a regression to data.

\boldmath
\subsubsection{$\eta\to\pi^0\gamma^\ast$}
\label{sec:Formfac_Eta}
\begin{figure}[t!]
    \centering
      \includegraphics[width=0.7\textwidth]{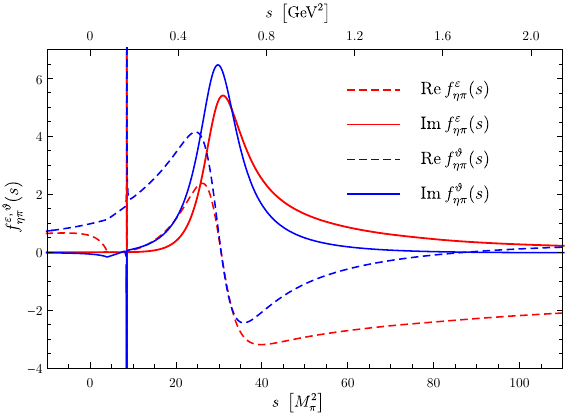}
      ~\\[.5cm]
      \includegraphics[width=0.7\textwidth]{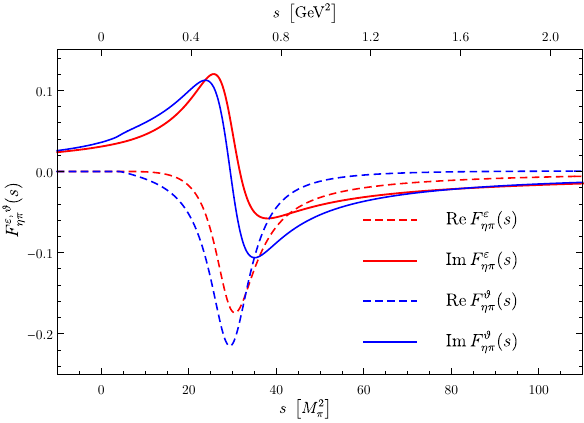}
    \caption{Basis solutions for the partial waves and form factors for the $\eta\to\pi^0$ transition. The partial-wave amplitudes $f_{\eta\pi}^\nu$ from Eq.~\eqref{eq:basis_solutions_partial_wave_EtaPi}
    are depicted in the upper panel; the singularity at the pseudothreshold $s=(M_\eta-M_{\pi^0})^2$ is clearly visible. These serve as an input to calculate the basis solutions of the transition form factors $F_{\eta\pi}^\nu(s)$ as defined in Eq.~\eqref{eq:BasisFormFac_Eta} and shown in the lower panel. 
    }
    \label{fig:BasisSolutions_Eta}
\end{figure}

\unboldmath
 The basis solutions for the $P$-wave amplitude $f\etapin$ are defined by
\beq\label{eq:basis_solutions_partial_wave_EtaPi}
    f^{\varepsilon}\etapin(s)\equiv\left.\big[\G_1(s)+\hat\G_1(s)\big]\right|_{\varepsilon=1}\,,   \qquad
    f^{\vartheta}\etapin(s)\equiv\left.\big[\H_1(s)+\hat\H_1(s)\big]\right|_{\vartheta=1}\,,
\eeq
and illustrated in Fig.~\ref{fig:BasisSolutions_Eta}(top). 
The dimensionless $f^{\varepsilon}\etapin$ corresponds to the isoscalar amplitude $\M_0^{\eta\pi}$, while $f^{\vartheta}\etapin$ belongs to the isotensor one, i.e., to $\M_2^{\eta\pi}$. The partial waves have a singular character at pseudothreshold, i.e., the upper limit in $s$ of the physical region in the $\eta\to \pi^0\pi^+\pi^-$ decay,
which is contained in the inhomogeneities describing left-hand-cut contributions to the respective partial wave.
Note that the form factor, after performing the dispersion integral over the discontinuity as in Eq.~\eqref{eq:isovector_formfactor}, is perfectly regular at that point.
Based on Eq.~\eqref{eq:basis_solutions_partial_wave_EtaPi}, we can calculate the corresponding \emph{basis form factors}  
\beq\label{eq:BasisFormFac_Eta}
    F^\nu_{\eta\pi}(s)=F^{(1)}_{\eta\pi}(s)|_{f_{\eta\pi}=f_{\eta\pi}^{\nu}} \eqwith \nu\in\{\varepsilon,\vartheta\}\,,
\eeq 
which allow us to linearly decompose $F\etapi^{(1)}$ according to
\beq\label{eq:FormFacDecomposition_Eta}
    F\etapin^{(1)}(s)=\varepsilon F\etapin^{\varepsilon}(s) + \vartheta F\etapin^{\vartheta}(s)\,.
\eeq
The $F\etapi^\nu$ are pure predictions of our dispersive representation, independent of the subtraction constants.
Our results for the basis solutions for the form factors are depicted in Fig.~\ref{fig:BasisSolutions_Eta}(bottom). 

Let us have a look at the hierarchy of the two amplitudes contributing to $F\etapin^{(1)}$. The plots in Fig.~\ref{fig:BasisSolutions_Eta} show that the basis solutions for the isoscalar and isotensor contributions to the $\eta\to \pi^0\pi^+\pi^-$ $P$-wave amplitude are of the same order of magnitude, and so are, as a result, the corresponding basis form factors. But due to the vast difference in their normalizing subtraction constants, the term $\vartheta F\etapin^{\vartheta}(s)$ is negligibly small in comparison to $\varepsilon F\etapin^{\varepsilon}(s)$. 
The origin of this discrepancy is well understood~\cite{Gardner:2019nid,Akdag:2021efj,Akdag:2023oob}.  The totally antisymmetric combination of $P$-wave single-variable functions in the isoscalar amplitude $\M^{\eta\pi}_0$, cf.\ Eq.~\eqref{eq:RT-eta3pi}, leads to a strong kinematic suppression inside the Dalitz plot; for symmetry reasons alone, the amplitude is required to vanish along the three lines $s=t$, $t=u$, and $u=s$.  As a result, the corresponding normalization $\varepsilon$ is far less rigorously constrained from fits to experimental data~\cite{Anastasi:2016cdz} than the isotensor amplitude, which only vanishes for $t=u$.  No such suppression occurs for the individual partial waves, or the transition form factors, be it in the $\rho$-resonance region or below, in the kinematic range relevant for the semi-leptonic decays studied here, where isoscalar and isotensor contributions show non-negligible, but moderate corrections to a $\rho$-dominance picture.  We also remark that this subtle interplay demonstrates that the model-independent connection between Dalitz plots and transition form factors absolutely requires the use of dispersion-theoretical methods---a low-energy effective theory such as chiral perturbation theory is insufficient for such extrapolations.

For the numerical evaluation of $F^{(1)}\etapi$ we only consider the by far dominant source of error, i.e., the uncertainty of the subtraction constants entering the partial wave. As their errors are of the same order of magnitude as their corresponding central values, it is a good approximation to neglect all other sources of uncertainties, such as the variation of phase-shift input.

\boldmath
\subsubsection{$\eta'\to\eta\gamma^\ast$}
\label{sec:Formfac_Etap}
\unboldmath

\begin{figure}[t!]
    \centering
      \includegraphics[width=0.7\textwidth]{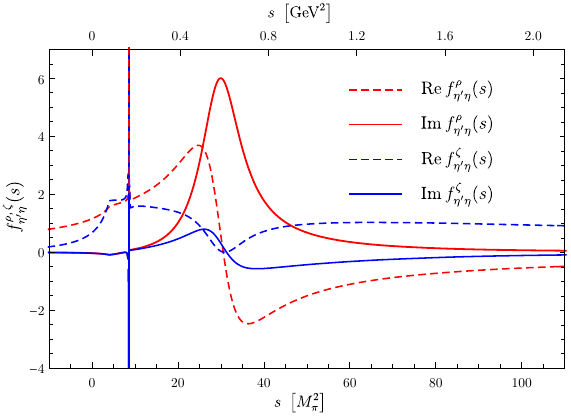}
      ~\\[.5cm]
      \includegraphics[width=0.7\textwidth]{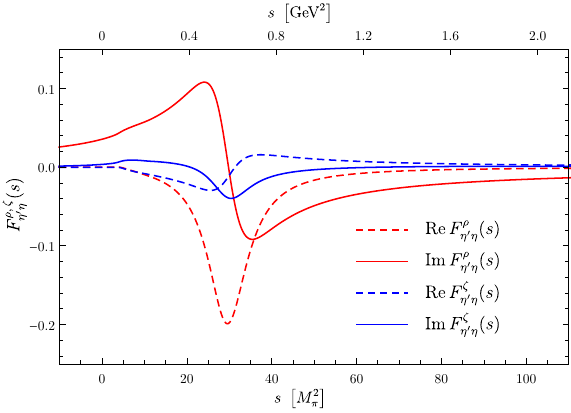}
    \caption{Basis solutions for the partial waves and form factors for the $\eta'\to\eta$ transition. The partial-wave amplitudes $f\etapeta^\nu$ from Refs.~\cite{Akdag:2021efj,Akdag:2023oob} 
    are depicted in the upper panel; again, the singularity at the pseudothreshold $s=(M_{\eta'}-M_{\eta})^2$ can be seen. These serve as an input to calculate the basis solutions of the transition form factors $F\etapeta^\nu(s)$ as defined in Eq.~\eqref{eq:BasisFormFac_Eta'} and shown in the lower panel. 
    }
    \label{fig:BasisSolutions_Etap}
\end{figure}

We now turn the focus on the transition form factor $f\etapeta$, whose basis solutions in terms of partial waves are defined as
\beq
    f^{\varrho}\etapeta(s)=\left.\big[\G\pipi(s)+\hat\G\pipi(s)\big]\right|_{\varrho=1,\,\zeta=0}\,, \qquad
    f^{\zeta}\etapeta(s)=\left.\big[\G\pipi(s)+\hat\G\pipi(s)\big]\right|_{\varrho=0,\,\zeta=1}\,.
\eeq
Using the $f^\nu\etapeta$ we can define the basis from factors
\beq\label{eq:BasisFormFac_Eta'}
    F^\nu\etapeta(s)=F^{(1)}\etapeta(s)|_{f\etapeta=f\etapeta^{\nu}} \eqwith \nu\in\{\varrho,\zeta\}\,,
\eeq 
and finally obtain the complete isovector form factor in explicit dependence on the subtraction constants by means of
\beq 
    F^{(1)}\etapeta(s)=\varrho F\etapeta^{\varrho}(s) + \zeta F\etapeta^{\zeta}(s)\,.
\eeq
The basis solutions for both partial waves and transition form factors are shown in Fig.~\ref{fig:BasisSolutions_Etap}.

\boldmath
\subsection{Resonance couplings from analytic continuation}
\label{sec:analytic_cont.}
\unboldmath

As both the partial waves $f_{XY}(s)$ and the resulting transition form factors $F^{(1)}_{XY}(s)$ have been constructed with the correct analytic properties, we can analytically continue them into the complex plane and onto the second Riemann sheet to extract resonance pole residues.  The resonance in question is the $\rho(770)$; its residues can be interpreted as model-independent definitions of $C$-violating $\rho\to XY$ coupling constants.
To this end, we recapitulate aspects of Refs.~\cite{Moussallam:2011zg, Hoferichter:2017ftn}. First, consider the discontinuity of the transition form factor in Eq.~\eqref{eq:disc} on the first Riemann sheet
\beq
    F^{(1),\,\text{I}}_{XY}(s+i\epsilon)- F^{(1),\,\text{I}}_{XY}(s-i\epsilon) = \frac{i}{24\pi}\big(\sigma^\pi(s+i\epsilon)\big)^3\,\big(F_\pi^{V,\,\text{I}}(s+i\epsilon)\big)^\ast\, \fI_{XY}(s+i\epsilon),
\eeq
with 
\beq
\sigma^\pi(s)\equiv \sqrt{\frac{4\mpi^2}{s}-1}\,, \qquad \sigma^\pi(s\pm i\epsilon)=\mp i\,\sigma_\pi(s)\,.
\eeq 
Using that the pion vector form factor fulfills Schwarz' reflection principle and demanding continuity of the scattering amplitudes when moving from one Riemann sheet to another, i.e.,
\beq
    F^{(1),\,\text{I}}_{XY}(s-i\epsilon)=F^{(1),\,\text{II}}_{XY}(s+i\epsilon) \eqand F_\pi^{V,\,\text{I}}(s-i\epsilon)=F_\pi^{V,\,\text{II}}(s+i\epsilon),
\eeq
we obtain
\beq\label{eq:PartialWaveSecondSheet}
    F^{(1),\,\text{II}}_{XY}(s+i\epsilon) = F^{(1),\,\text{I}}_{XY}(s+i\epsilon)-\frac{i}{24\pi}\big(\sigma^\pi(s+i\epsilon)\big)^3\,F_\pi^{V,\,\text{II}}(s+i\epsilon)\,f^\text{I}_{XY}(s+i\epsilon)\,. 
\eeq
Left- and right-hand side of Eq.~\eqref{eq:PartialWaveSecondSheet} depend on the same argument, such that, by analytic continuation, this relation can be applied in the whole complex plane.  In particular, in the vicinity of the $\rho(770)$ pole, the transition form factors as well as the pion form factor on the second Riemann sheet behave as 
\beq
   F^{(1),\,\text{II}}_{XY}(s)\,,\ F_\pi^{V,\,\text{II}}(s)\propto\frac{1}{s_\rho-s} \eqwith 
s_\rho=\left( M_\rho-i\frac{\Gamma_\rho}{2}\right)^2.
\eeq
The pole position $s_\rho$ 
has been determined most accurately in Ref.~\cite{Garcia-Martin:2011nna}, using Roy-like equations for pion--pion scattering: 
$M_\rho=763.7\MeV$, $\Gamma_\rho=146.4\MeV$ (cf.\ also Ref.~\cite{Colangelo:2001df}); for later use, we also quote the coupling constant to $\pi\pi$, $|g_{\rho\pi\pi}|=6.01$, $\arg(g_{\rho\pi\pi}) = -5.3^\circ$.  We neglect the uncertainties in these parameters, as they are small compared to the ones fixing the partial waves $f_{XY}$.
While $F_\pi^{V,\,\text{II}}$ is explicitly given in Ref.~\cite{Hoferichter:2017ftn}, we can match $F^{(1),\,\text{II}}_{XY}$ to a VMD-type form factor similar to Eq.~\eqref{eq:FormFacVMD}, but with $g_{\rho YX}$, $g_{\rho\gamma}$, and $M^2_\rho$ instead of $\tilde g$, $g_{v\gamma}$, and $M^2_v$. Thus, in sufficient vicinity to the pole, we can write 
\beq\label{eq:PWA_second_sheet}
	F_\pi^{V,\,\text{II}}(s)= \frac{g_{\rho\pi\pi}}{g_{\rho\gamma}}\frac{s_\rho}{s_\rho-s}
    \eqand
    F^{(1),\,\text{II}}_{XY}(s)= \frac{g_{\rho YX}}{2g_{\rho\gamma}}\frac{M^2_\rho}{s_\rho-s}.
\eeq
If we evaluate Eq.~\eqref{eq:PartialWaveSecondSheet} near the pole $s_\rho$  and insert Eq.~\eqref{eq:PWA_second_sheet}, we can compute the desired $C$-odd $\rho$-meson couplings by 
\begin{aligneq}\label{eq:vector_meson_couplings}
    g_{\rho YX}&
    =\frac{g_{\rho\pi\pi}}{12\pi}\frac{s_\rho}{M^2_\rho}\sigma^3_\pi(s_\rho)\,\fI_{XY}(s_\rho)\,.
\end{aligneq}
The problem is therefore reduced to evaluating the partial wave $\fI_{XY}$ on the first Riemann sheet at the pole position, a task for which the dispersive representations are perfectly suited. To clarify the dependence on subtractions or effective coupling constants and therefore separate the uncertainty in these from the precisely calculable dispersive aspects, we will once more make use of the decomposition in terms of basis functions.

We begin with the $\eta\to\pi^0$ transition form factor.
The basis functions of the partial wave $f\etapin$, evaluated at the $\rho$ pole, result in 
\beq
f\etapin^\varepsilon(s_\rho)=-0.02 - 2.76i \,, \qquad
f\etapin^\vartheta(s_\rho)=0.87 -3.05i \,,
\eeq
so that we obtain
\begin{equation}\label{eq:analytic_con_eta}
f\etapin(s_\rho)= \varepsilon f\etapin^\varepsilon(s_\rho)+\vartheta f\etapin^\vartheta(s_\rho)=\big[(-0.704+i\,0.005)\,\text{GeV}^4 \, g_0-(1.149+i\,0.330)g_2\big] \,,
\end{equation}
where we made use of Eq.~\eqref{eq:subcons_couplings_eta}.
Employing of Eq.~\eqref{eq:vector_meson_couplings} and finally inserting the values for the coupling constants $g_0$ and $g_2$ as extracted from the $\eta\to\pi^0\pi^+\pi^-$ Dalitz-plot asymmetry then yields
\begin{aligneq}
    g_{\rho\pi\eta}
    &= \big[(-0.089 + i\,0.022)\,\text{GeV}^4 \, g_0 -(0.156 + i\,0.007) \, g_2\big] \\[0.1cm]
   &=\big[0.25(0.40)-i\,0.06(0.10)\big]\,\mathrm{GeV}^{-2} \,.
   \label{eq:g_rho-pi-eta}
\end{aligneq}
Note that the isotensor contribution $g_2$ is negligible in the coupling $g_{\rho\pi\eta}$.

The analytic continuation of the basis partial wave for $f\etapeta$ to the pole position of the $\rho$ meson yields 
\beq
f\etapeta^\varrho(s_\rho)=0.44 - 2.95 i \,, \qquad
f\etapeta^\zeta(s_\rho)=-0.45 - 0.17i\,.
\eeq
With Eq.~\eqref{eq:subcons_couplings_eta'} we can hence express the analytically continued partial wave at the $\rho$ pole by
\begin{equation}\label{eq:analytic_con_eta'}
f\etapeta(s_\rho)= \varrho f\etapeta^\varrho(s_\rho)+\zeta f\etapeta^\zeta(s_\rho)=g_1\left[(0.12-i\,0.34)+(1.63+i\,0.48)\,\text{GeV}^{2}\, \delta g_1\right]\,.
\end{equation} 
In terms of this result, the $\rho$-meson coupling from Eq.~\eqref{eq:vector_meson_couplings} results in
\begin{aligneq}
 g_{\rho\eta\eta'}
 &=g_1\left[(0.005-i\,0.047)+(0.222+i\,0.011)\,\text{GeV}^{2}\, \delta g_1\right] \\[0.1cm]
&= -\big[0.85(2.36)+i\,0.08(0.17)\big]\,\mathrm{GeV}^{-2} 
\label{eq:g_rho-eta-eta'}\,,
\end{aligneq}
where we considered correlated Gaussian errors for the couplings $g_1$ and $\delta g_1$.

Note that the coupling constants in Eq.~\eqref{eq:vector_meson_couplings} become inevitably complex-valued, thus spoiling the well-defined transformation under time reversal when compared to the tree-level coupling constants from ToPe$\chi$PT. This is neither surprising nor specific to the context of symmetry violation studied here: in the strong interactions, resonance couplings that are real in the narrow-width limit necessarily turn complex when defined model-independently via pole residues in the complex plane. 
However, this points towards the reason why these complex phases will be irrelevant when using symmetry arguments to estimate isoscalar contributions in the next section: for the narrow $\omega$ and $\phi$ resonances, they are negligible to far better accuracy; symmetry arguments within the vector-meson nonet are not applicable to their total widths.  We will therefore simply omit the imaginary parts in the next section and relate the $C$-odd $\omega$ couplings required for the model of the isoscalar parts of the form factors to the real parts of the $\rho$ coupling (of the same total isospin) only.  Note furthermore that Eqs.~\eqref{eq:g_rho-pi-eta} and \eqref{eq:g_rho-eta-eta'} still suggest the imaginary parts of $g_{\rho\pi\eta}$ and $g_{\rho\eta\eta'}$ to be rather small, such that the difference between real part and modulus, e.g., is negligible for our purposes.

\boldmath
\section{Hadronic long-range effects: the isoscalar contribution}
\label{sec:Hadronic_long_range_isoscalar}
\unboldmath
We now attempt to combine the findings of Sects.~\ref{sec:isoscalar} and~\ref{sec:analytic_cont.}.  
We wish to access the couplings $\tilde g$, cf.\ Eq.~\eqref{eq:FormFacVMD}, by linking them to the $g_{\rho YX}$ discussed in the last section. 
In Sect.~\ref{sec:isoscalar} we found a ToPe$\chi$PT operator that, when considered separately, allows us---according to Eq.~\eqref{eq:correlated_couplings}---to relate these couplings by $SU(3)$ symmetry. The vector-meson couplings with the same total isospin are found to be related by $g_{\omega\pi\eta}=1/\sqrt{3}\, g_{\rho\eta\eta'}$ and 
$g_{\omega\eta\eta'}= \sqrt{3} \,g_{\rho\pi\eta}$, 
while $g_{\phi\pi\eta}=0$ and $g_{\phi\eta\eta'}$ does \emph{not} correlate with respective $\rho$ couplings. However, the predictive power of flavor symmetry arguments does not hold in general for all operators.
This leads to the shortcoming that we cannot fix the relative sign of the couplings, which becomes evident when comparing Eqs.~\eqref{eq:uncorrelated_couplings} and \eqref{eq:correlated_couplings}, and have to rely on NDA arguments to consider that there may be additional contributions to the couplings from linear combinations of Wilson coefficients, cf.\ Eq.~\eqref{eq:uncorrelated_couplings}. An alternative approach would be to use NDA right away and drop the relative factors of $1/\sqrt{3}$ and $\sqrt{3}$, respectively, but this still leads to the same caveats.

\boldmath
\subsection{$\eta\to\pi\ell^+\ell^-$}
\label{sec:isoscalar_Eta}
\unboldmath

Possible contributions to the isoscalar form factor in $\eta\to\pi^0\ell^+\ell^-$ can originate from an $\omega$ or a $\phi$ intermediate state. In accordance with Sect.~\ref{sec:isoscalar}, these enter the form factor in the linear combination 
\beq\label{eq:isoscalar_formfac_eta}
    F_{\eta\pi}^{(0)}(s)\equiv
    \frac{g_{\omega\pi\eta}}{2g_{\omega\gamma}}\frac{M^2_\omega}{M^2_\omega-s}+\frac{g_{\phi\pi\eta}}{2g_{\phi\gamma}}\frac{M^2_\phi}{M^2_\phi-s}\,.
\eeq
With our $SU(3)$ estimate $g_{\phi\pi\eta}=0$ we can ignore the contribution of the $\phi$. Dropping the latter is also justified from an NDA point of view:  
the difference of the two summands in Eq.~\eqref{eq:isoscalar_formfac_eta} is negligible compared to the uncertainty of NDA if $F_{\eta\pi}^{(0)}(s)$ is evaluated within the physical range. Therefore, we continue the estimation of the isoscalar contribution with the $\omega$ intermediate state only, for which we use $|g_{\omega\gamma}|=16.7(2)$~\cite{Hoferichter:2017ftn}.

Relating $g_{\omega\pi\eta}$ to the $\rho$ coupling of the same total isospin $I=1$ and omitting the imaginary part for the reasons given above, we find
\beq\label{eq:g-omega-pi-eta-isovector} 
g_{\omega\pi\eta}\approx \frac{1}{\sqrt{3}}\Re g_{\rho\eta\eta'}
= -0.49(1.36)\,\mathrm{GeV}^{-2} \eqor  |g_{\omega\pi\eta}|\lesssim 1.9\,\mathrm{GeV}^{-2}\,.
\eeq

Throughout this manuscript we do not account for the numerically intangible uncertainties from our $SU(3)$ estimates or NDA. As neither of the latter fixes the sign of $|g_{\omega\pi\eta}|$, we have to content ourselves with its absolute value.

Note that retaining the imaginary part of $g_{\rho\eta\eta'}$ would have a negligible effect on the upper limit for $|g_{\omega\pi\eta}|$.

On the other hand, we can also place a bound on 
$|g_{\omega\pi\eta}|$ using the upper limit on the branching ratio of $\omega\to\eta\pi^0$ as determined by the Crystal Ball multiphoton spectrometer at the Mainz Microtron (MAMI)~\cite{Starostin:2009zz} and the Lagrangian in Eq.~\eqref{eq:generic_vector_meson_Lagrangian}. 
The partial decay width is found to be 
\beq
\Gamma(\omega\rightarrow\eta\pi^0)= 
\frac{1}{192\pi M_\omega}\, 
    |g_{\omega\pi\eta}|^2\,
    \lambda^{3/2}\big(M_\omega^2,\meta^2,\mpii^2\big)\,.
\eeq
With $\text{BR}(\omega\rightarrow\eta\pi^0)<2.3\cdot 10^{-4}$~\cite{Starostin:2009zz} and $\Gamma_\omega= 8.68$~MeV~\cite{Workman:2022ynf}, we obtain the bound
\beq\label{eq:coupling_limit_MAMI}
  |g_{\omega\pi\eta}| <0.24\,\mathrm{GeV}^{-2}\,,
\eeq
which is significantly more restrictive than the theoretical estimate for the bound on the coupling inferred from $g_{\rho\eta\eta'}$.

\boldmath
\subsection{$\eta'\to\eta\ell^+\ell^-$}
\label{sec:isoscalar_Etap}
\unboldmath

Similarly to the previous section, the isoscalar part of the form factor in $\eta'\to\eta\ell^+\ell^-$ can be written as
\beq\label{eq:isoscalar_formfac_eta'}
    F\etapeta^{(0)}(s)\equiv
    \frac{g_{\omega\eta\eta'}}{2g_{\omega\gamma}}\frac{M^2_\omega}{M^2_\omega-s}+\frac{g_{\phi\eta\eta'}}{2g_{\phi\gamma}}\frac{M^2_\phi}{M^2_\phi-s}\,.
\eeq
With the same reasoning as above we henceforth drop the contribution of the $\phi$ and only take the $\omega$ into account.
The numerical result for the corresponding vector meson coupling, which has total isospin $I=0$, is 
\beq\label{eq:upper_limit_vector_coupling_Eta'}
    g_{\omega\eta\eta'} \approx \sqrt{3}\, \Re g_{\rho\pi\eta} = 0.43(0.69)\,\mathrm{GeV}^{-2} \eqor
    |g_{\omega\eta\eta'}|\lesssim 1.1\,\mathrm{GeV}^{-2}\,.
\eeq

Once more, the imaginary part of $g_{\rho\pi\eta}$ would yield just a minor contribution to the upper limit on $|g_{\omega\eta\eta'}|$ and can be neglected. We furthermore remark that the $\rho\pi\eta$ coupling also has an isotensor component, which, however, has a negligible effect, cf.\ Eq.~\eqref{eq:g_rho-pi-eta}.

\section{Results}
\label{sec:results}
With the theoretical apparatus at hand we are now able to predict upper limits on the decay widths 
\beq\label{eq:decay_width_results}
    \Gamma(X\to Y \ell^+ \ell^-)
    =\frac{\alpha^2}{8\pi M_X^3}\,\int_{4 m_\ell^2}^{(M_X-M_Y)^2} \!\diff s\, \lambda^{3/2}\big(s,M_X^2,M_Y^2\big)\, \sigma_\ell(s)\left(1-\frac{\sigma^2_\ell(s)}{3}\right) |F_{XY}(s)|^2\,,
\eeq
relying on the Dalitz-plot asymmetries in $X\to Y \pi^+\pi^-$ as the main input. As argued in Sect.~\ref{sec:discussion_phenomenology}, we focus on the long-range contributions via hadronic intermediate states only, i.e., we set
\beq
    F_{XY}(s)=F^{(1)}_{XY}(s)+F^{(0)}_{XY}(s)\,.
\eeq
We disregard the contributions analyzed in Sects.~\ref{sec:Direct_Semi-Leptonic} and \ref{sec:Direct_Photonic} according to the discussion in Sect.~\ref{sec:discussion_phenomenology}: these do not show interesting correlations with other ToPe processes, and absent significant cancellations, we can study the consequences of limit setting for the long-range hadronic effects alone.
The corresponding transition form factors for the isovector and isoscalar contributions to $\eta\to\pi^0\gamma^\ast$ read
\begin{aligneq}
   F\etapin^{(1)}(s)=\varepsilon F\etapin^{\varepsilon}(s) + \vartheta F\etapin^{\vartheta}(s)\,,\qquad
    F_{\eta\pi}^{(0)}(s)=
    \frac{g_{\omega\pi\eta}}{2g_{\omega\gamma}}\frac{M_\omega^2}{M^2_\omega-s}\,,
\end{aligneq}
while the ones contributing to $\eta'\to\eta\gamma^\ast$ are
\begin{aligneq}
     F^{(1)}\etapeta(s)=\varrho F\etapeta^{\varrho}(s) + \zeta F\etapeta^{\zeta}(s)\,,\qquad 
     F^{(0)}_{\eta'\eta}(s)=
    \frac{g_{\omega \eta\eta'}}{2g_{\omega\gamma}}\frac{M^2_\omega}{M^2_\omega-s}\,.
\end{aligneq}
The subtraction constants fixing the $F^{(1)}_{XY}$ are given in Eqs.~\eqref{eq:subcons_eta} and \eqref{eq:subcons_eta'}, the respective basis solutions $F_{XY}^{\nu}$ are depicted in Figs.~\ref{fig:BasisSolutions_Eta} and~\ref{fig:BasisSolutions_Etap}, and the coupling constants $g_{\omega YX}$ entering the $F^{(0)}_{XY}$ are quoted in Eqs.~\eqref{eq:coupling_limit_MAMI} and \eqref{eq:upper_limit_vector_coupling_Eta'}, respectively.

We have pointed out above that we have no means to assess the relative sign of the isoscalar contribution.  To determine theoretical upper bounds, we 
investigate the parameter space spanned by the coupling constants and their uncertainties fixing 
\beq\label{eq:AbsoluteValue}
    |F\etapin|^2
    =
    |F\etapin^{(0)}|^2+|F\etapin^{(1)}|^2+2\,\Re \big(F\etapin^{(0)}\,F\etapin^{(1)\ast}\big)
\eeq
and similarly for $F\etapeta$.

\boldmath
\subsection{$\eta\to\pi\ell^+\ell^-$}
\label{sec:results_eta}
\unboldmath

\begin{figure}[t!]
    \centering
      \includegraphics[width=0.7\textwidth]{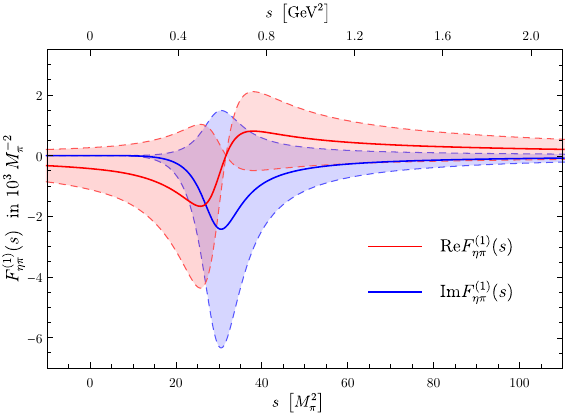}
      ~\\[.5cm]
      \includegraphics[width=0.7\textwidth]{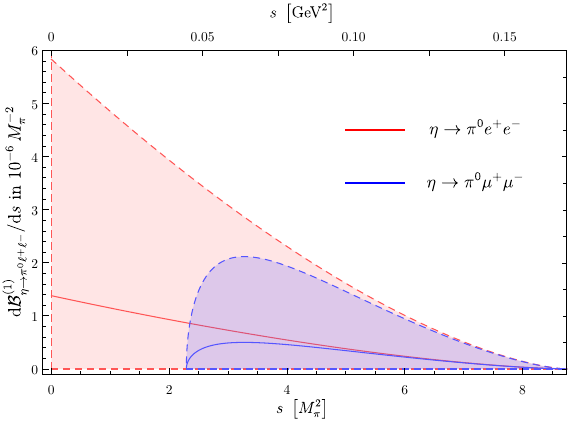}
    \caption{Spectrum of the isovector contribution to the form factor (top) and the corresponding differential decay distribution (bottom) for $\eta\to \pi^0 \ell^+\ell^-$. The dashed lines mark the respective upper and lower limits stemming from the uncertainties of the subtraction constants in Eq.~\eqref{eq:subcons_eta}. The physical ranges are in both cases restricted by $4m_\ell^2\leq s\leq (\meta-\mpi)^2$.
    }
    \label{fig:dGamma/ds_eta}
\end{figure}

Based on the splitting of the absolute square of the transition form factor
given in Eq.~\eqref{eq:AbsoluteValue},
the branching ratios $\B_{\eta\to\pi^0 e^+ e^-}$ and $\B_{\eta\to\pi^0 \mu^+ \mu^-}$
can be expressed in terms of the contributing BSM coupling constants $g_0$, $g_2$,
and $g_{\omega\pi\eta}$ in the following way:
\begin{aligneq}\label{eq:WidthsEtaPicoeff}
    \B_{\eta\to\pi^0 e^+ e^-}& =10^{-6}\Big(&&
    0.61 \, \mathrm{GeV}^{12} \, g_0^2
    +2.20 \, \mathrm{GeV}^{8} \, g_0\,g_2
    +1.99 \, \mathrm{GeV}^{4} \, g_2^2\\
    &&&
    +4.68 \, \mathrm{GeV}^{8} \, g_0\,g_{\omega\pi\eta}
    +8.44\, \mathrm{GeV}^{4} \, g_2\,g_{\omega\pi\eta}
    +8.97 \, \mathrm{GeV}^{4} \, g^2_{\omega\pi\eta} \Big) \,,\\[0.2cm]
    \B_{\eta\to\pi^0 \mu^+ \mu^-}&
    =10^{-6}\Big(&&
    0.22 \, \mathrm{GeV}^{12} \, g_0^2
    +0.81 \, \mathrm{GeV}^{8} \, g_0\,g_2
    +0.76\, \mathrm{GeV}^{4} \, g_2^2\\
    &&&
    +1.66 \, \mathrm{GeV}^{8} \, g_0\,g_{\omega\pi\eta} 
    +3.10\, \mathrm{GeV}^{4} \, g_2\,g_{\omega\pi\eta}
    +3.18 \, \mathrm{GeV}^{4} \, g^2_{\omega\pi\eta}\Big)\,.
\end{aligneq}
If we insert the maximum-range values $g_0 = (-2.8-4.5)\,\text{GeV}^6$, $g_2= (-9.3-4.6)\cdot 10^{-3}\,\text{GeV}^{-2}$,  and $|g_{\omega\pi\eta}| = 0.24\,\text{GeV}^{-2} $, cf.\ Sect.~\ref{sec:Eta3Pi_partial_wave} and Eq.~\eqref{eq:coupling_limit_MAMI}, in Eq.~\eqref{eq:WidthsEtaPicoeff}, 
we obtain the following conservative limits:
\begin{align}\label{eq:WidthsEtaPi}
        \B_{\eta\to\pi^0 e^+ e^-}^{(1)}&<
        33\cdot 10^{-6}, 
        &  
        \B_{\eta\to\pi^0 e^+ e^-}&<
        41\cdot 10^{-6}, 
        &  
        \B_{\eta\to\pi^0 e^+ e^-}^\text{exp} &< 7.5\cdot 10^{-6} \,,
        \notag \\[0.2cm]
        \B_{\eta\to\pi^0 \mu^+ \mu^-}^{(1)}&<
        12\cdot 10^{-6}, 
        &  
        \B_{\eta\to\pi^0 \mu^+ \mu^-}&<
        15\cdot 10^{-6},
        &  
        \B_{\eta\to\pi^0 \mu^+ \mu^-}^\text{exp} &< 5\cdot 10^{-6} \,.
\end{align}
Here the first entry in each line corresponds to the isovector contribution and the second includes the isoscalar one in addition. 
Finally, the experimental results~\cite{WASA-at-COSY:2018jdv,Dzhelyadin:1980ti}, to be understood at 90\% C.L., are quoted last (cf.\ Table~\ref{tab:BR+SM}), which 
are smaller than our findings by a factor $5.5$ and $3$, respectively.
The isovector contributions to the respective form factors and differential decay widths for the decay $\eta\to\pi\ell^+\ell^-$ are shown in Fig.~\ref{fig:dGamma/ds_eta}.
If the area enclosed by the error bands of the differential decay distribution is integrated, compatible but somewhat less conservative limit values result.

Assuming that the other two coupling constants are negligible, the relations~\eqref{eq:WidthsEtaPicoeff} can be used to derive upper bounds for, respectively, one of the three BSM coupling constants from the experimental limits of the branching ratios listed in Table~\ref{tab:BR+SM}. In the same table, the Standard-Model predictions for the dilepton branching ratios based on the $C$-even two-photon mechanism~\cite{Schafer:2023qtl} can be found. These can be used to calculate the lower threshold for the aforementioned coupling constants, at which the BSM scenario would no longer dominate the Standard-Model case. The upper bounds and the lower thresholds, which we refer to as sensitivities for the sake of simplicity, are listed in the third and fourth columns of Table~\ref{tab:BR_eta}, respectively. 
\begin{table}
\centering
\renewcommand{\arraystretch}{1.3}
\begin{tabular}{lrcc}
\toprule
BSM/hadronic & Empirical values via & Limits derived in \eqref{eq:WidthsEtaPicoeff} 
& Sensitivities in \eqref{eq:WidthsEtaPicoeff} \\
coupl.\,constants     &hadronic constraints & 
from exp.\ bounds& from SM predictions  \\   
\midrule
$g_0~~~~~\,[\text{GeV}^{-6}]$ &$-2.8(4.5)$~\cite{Akdag:2021efj}  & (3.5 | 4.8)& $( 5.0 \ | \ 5.8) \cdot 10^{-2}$  \\
$g_2~~~~~\,[\text{GeV}^{-2}]$ &$-9.3(4.6)\cdot 10^{-3}$~\cite{Akdag:2021efj}  & (1.9 | 2.6) & $( 2.8 \ | \ 3.1) \cdot 10^{-2}$\\
$|g_{\omega\pi\eta}|~[\text{GeV}^{-2}] $  & 0.24~\cite{Starostin:2009zz}
 & (0.9 | 1.3) & $(1.3 \ | \  1.5)\cdot 10^{-2}$\\
\bottomrule
\end{tabular}
\renewcommand{\arraystretch}{1.0}
\caption{Empirical values, upper limits, and  lower thresholds (sensitivities) of the BSM  hadronic coupling constants $g_0$, $g_2$, and (the modulus of) $g_{\omega\pi\eta}$, derived  (i) from the $\eta\to \pi^0\pi^+\pi^-$  Dalitz-plot asymmetries of 
Ref.~\cite{Akdag:2021efj} and the empirical partial width of the decay 
$\omega\to \eta\pi^0$~\cite{Starostin:2009zz}, (ii) from the experimental upper bounds on the  branching ratios $\B_{\eta\to\pi^0 \ell^+ \ell^-}$ given in
Table~\ref{tab:BR+SM} for the dilepton pairs  ($e^+e^-$ | $\mu^+\mu^-$), as calculated with aid of  the relations~\eqref{eq:WidthsEtaPicoeff}
assuming that the other two coupling constants are vanishingly small, and (iii) as in (ii), but from the SM predictions listed in Table~\ref{tab:BR+SM}.
\label{tab:BR_eta}}
\end{table}
The values to the left of the vertical bar refer to the case of $e^+ e^-$ dileptons, while the values to the right describe the results of the $\mu^+\mu^-$ case. 
In addition, we also list the empirical values of $g_0$ and $g_2$, derived via the $\eta\to \pi^0\pi^+\pi^-$ Dalitz-plot asymmetries in Ref.~\cite{Akdag:2021efj}, and the upper experimental bound~\eqref{eq:coupling_limit_MAMI} of $|g_{\omega\pi\eta}$|, derived from the partial decay width
$\Gamma(\omega\to \eta\pi^0)$~\cite{Starostin:2009zz}, in the second column of Table~\ref{tab:BR_eta}. 

The empirical hadronic values of $g_0$ and $g_{\omega\pi\eta}$ are approximately of the same order of magnitude as the upper bounds calculated according to Eq.~\eqref{eq:WidthsEtaPicoeff}.
In contrast, the upper bound  from Eq.~\eqref{eq:WidthsEtaPicoeff} is not competitive at all
in the case of $g_2$, since it is more than two orders of magnitude larger than the 
(modulus of the) empirical value. Moreover, the latter is even smaller than the SM sensitivity for both decays $\eta\to\pi^0\ell^+\ell^-$, $\ell =e,\mu$. Therefore, {these} decays cannot be used to {further} constrain this parameter of $C$-odd, $P$-even $\eta$ decays.

When making such comparisons, it must be kept in mind that the calculations of the upper bounds and lower threshold values are based on the strict assumption that the other two couplings are negligible. Should one of these three coupling constants be different from zero, it would be very unlikely that none of the other two should play any role.

In summary, the considered experiments for $\eta\to\pi^0\pi^+\pi^-$ and $\eta\to\pi^0\ell^+\ell^-$ have a quite similar sensitivity for ToPe forces of total isospin 0 as exemplified by the coupling $g_0$, despite the fact that asymmetries in the former are based on $C$-odd interferences and therefore scale \emph{linearly} with a (small) BSM coupling~\cite{Gardner:2019nid, Akdag:2022sbn}, while the latter is a rate that is suppressed to second order in a similar coupling. 
This is not so much due to a high sensitivity of the semi-leptonic decays, but rather to the strong suppression of $g_0$ in the $\eta\to\pi^0\pi^+\pi^-$ Dalitz plot.
Comparing $\B_{\eta\to\pi^0 \ell^+ \ell^-}^{(1)}$ to $\B_{\eta\to\pi^0 \ell^+ \ell^-}$, our analysis suggests that the isoscalar form factor contributes roughly one quarter to the overall branching ratio. This reflects the more rigorous bound on the $\omega\to\eta\pi^0$ coupling, cf.\ Eq.~\eqref{eq:coupling_limit_MAMI}, compared to the limits inferred from the $\eta\to\pi^0\pi^+\pi^-$ Dalitz-plot asymmetries for the isovector part of the form factor.

\begin{sloppypar}
As the experimental limits turn out to be more restrictive than our theoretical predictions for $\B_{\eta\to\pi^0 \ell^+ \ell^-}^{(1)}$, the empirical $\eta\to \pi^0\ell^+\ell^-$ decay widths~\cite{WASA-at-COSY:2018jdv, Dzhelyadin:1980ti} can be used to refine the fit to the $\eta\to\pi^0\pi^+\pi^-$ Dalitz plot~\cite{Anastasi:2016cdz}. As long as the latter constrains the corresponding BSM couplings in a way that the form factor is dominated by the contribution of $g_0$, an improved regression to the full Dalitz plot is redundant. Instead we note that isoscalar ($g_0$) and isotensor ($g_2$) couplings in $\eta\to\pi^0\pi^+\pi^-$ are nearly uncorrelated~\cite{Akdag:2021efj} and turn the experimental  limit for $\B_{\eta\to\pi^0 e^+ e^-}^{(1)}$ 
into the upper bound in  Table~\ref{tab:BR_eta},
 \beq\label{eq:emp_bound_g0}
|g_0|<3.5\,\text{GeV}^{-6}\,,
 \eeq
to be compared to the previous constraint $g_0=-2.8(4.5)\,\text{GeV}^{-6}$~\cite{Akdag:2021efj}.
\end{sloppypar}

\boldmath
\subsection{$\eta'\to\eta\ell^+\ell^-$}
\label{sec:results_etap}
\begin{figure}[t!]
    \centering
      \includegraphics[width=0.7\textwidth]{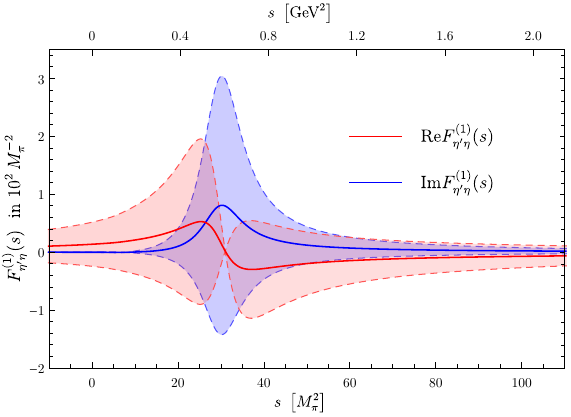}
      ~\\[.5cm]
      \includegraphics[width=0.7\textwidth]{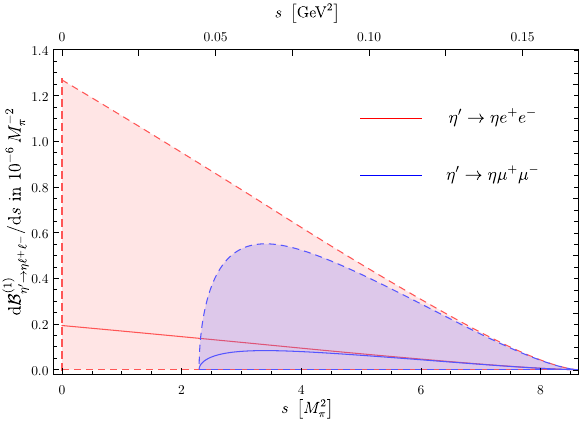}
    \caption{Spectrum of the isovector contribution to the form factor (top) and the corresponding  differential decay distribution (bottom) for $\eta'\to \eta \ell^+\ell^-$. The dashed lines mark the respective upper and lower limits stemming from the uncertainties of the subtraction constants in Eq.~\eqref{eq:subcons_eta'}. The physical ranges are in both cases restricted by $4m_\ell^2\leq s\leq (\metap-\meta)^2$.}
    \label{fig:dGamma/ds_eta'}
\end{figure}

\unboldmath
Proceeding in analogy to Sect.~\ref{sec:results_eta}, we obtain the following 
relations for the branching ratios 
in explicit dependence on the BSM couplings $g_1$, $\delta g_1$, and $g_{\omega\eta\eta'}$:
\begin{aligneq}\label{eq:WidthsEtapEtacoeff}
    \B_{\eta'\to\eta e^+ e^-}&=10^{-8}\Big(&&
    0.23\, \mathrm{GeV}^{4} \,g_1^2
    +2.14\, \mathrm{GeV}^{6} \, g_1^2 \,\delta g_1
    +5.66\, \mathrm{GeV}^{8} \, g_1^2 \,\delta g_1^2
    \\
    &&&
    -3.35 \, \mathrm{GeV}^{4} \,g_1\,g_{\omega\eta\eta'}
    -17.76 \, \mathrm{GeV}^{6} \, g_1\delta g_1 \,g_{\omega\eta\eta'}
    +13.95\, \mathrm{GeV}^{4} \, g_{\omega\eta\eta'}^2
    \Big) \,,\\[0.2cm]
    \B_{\eta'\to\eta \mu^+ \mu^-}&=10^{-8}\Big(&&
    0.13\, \mathrm{GeV}^{4} \, g_1^2
    +1.05\, \mathrm{GeV}^{6} \, g_1^2 \,\delta g_1
    +2.25\, \mathrm{GeV}^{8} \, g_1^2 \, \delta g_1^2
    \\
    &&&
    -1.62 \, \mathrm{GeV}^{4} \, g_1 \, g_{\omega\eta\eta'}
    -6.98\, \mathrm{GeV}^{6} \, g_1 \, \delta g_1 \, g_{\omega\eta\eta'} 
    +5.43\, \mathrm{GeV}^{4} \, g_{\omega\eta\eta'}^2
    \Big)\,.
\end{aligneq}
If the maximum-range values $g_1 = (0.7+1.0)\,\text{GeV}^{-2}$, $g_1\delta g_1 = (-3.9-10.5)\,\text{GeV}^{-4}$ (interpreted as one effective coupling constant),\footnote{Note that the errors on $g_1$ and $g_1\delta g_1$ are almost perfectly anticorrelated~\cite{Akdag:2021efj,Akdag:2023oob}.} and $g_{\omega\eta\eta'} = (0.43+0.69)\,\text{GeV}^{-2}$, cf.\ Sect.~\ref{sec:EtapEtaPiPi_partial_wave} and Eq.~\eqref{eq:upper_limit_vector_coupling_Eta'}, are inserted in Eq.~\eqref{eq:WidthsEtapEtacoeff}, the conservative estimates of the upper limits for the decays $\eta'\to\eta\ell^+\ell^-$ are\footnote{For these branching ratios we use the total width $\Gamma_{\eta'}=0.23\,\text{MeV}$ listed as \emph{PDG average} in Ref.~\cite{Workman:2022ynf}. Again the experimental branching ratios~\cite{CLEO:1999nsy,Dzhelyadin:1980ti} are to be understood at 90\% C.L., cf.\ Table~\ref{tab:BR+SM}, and the isovector part is designated with the upper index $(1)$.}
\begin{align}\label{eq:WidthsEtapEta}
        \B_{\eta'\to\eta e^+ e^-}^{(1)}&<
        11\cdot 10^{-6}, 
        & 
        \B_{\eta'\to\eta e^+ e^-}&<
        14\cdot 10^{-6}, 
        & 
        \B_{\eta'\to\eta e^+ e^-}^\text{exp} &< 2.4\cdot 10^{-3} \,,
        \notag \\[0.2cm]
        \B_{\eta'\to\eta \mu^+ \mu^-}^{(1)}&<
        4.4\cdot 10^{-6}, &
        \B_{\eta'\to\eta \mu^+ \mu^-}&<
        5.6\cdot 10^{-6}, &
        \B_{\eta'\to\eta \mu^+ \mu^-}^\text{exp} &< 15\cdot 
        10^{-6} 
        \,.
\end{align}
For these decays, our approximation for the isoscalar form factor loosens the limit on the isovector part by roughly a quarter. 
A depiction of the latter contribution to the form factor and differential decay width is given in Fig.~\ref{fig:dGamma/ds_eta'}.
In contrast to the findings for $\B_{\eta\to\pi^0 \ell^+ \ell^-}$, our limits 
on $\B_{\eta'\to\eta \ell^+ \ell^-}$ are more restrictive than the respective 
experimental ones.

\begin{table}
\centering
\renewcommand{\arraystretch}{1.3}
\begin{tabular}{lrcc}
\toprule
BSM/hadronic & Empirical values via & Limits derived in \eqref{eq:WidthsEtapEtacoeff} 
& Sensitivities in \eqref{eq:WidthsEtapEtacoeff} \\
coupl.\,constants     &hadronic constraints & 
from exp.\ bounds& from SM predictions  \\   
\midrule
$g_1~~~~~[\text{GeV}^{-2}]$ &$0.7(1.0)$~\cite{Akdag:2023oob}  & $(1000  | \ 110) $& $( 0.49 \ | \ 0.45) $  \\
$g_1\delta g_1~[\text{GeV}^{-4}]$ &$-3.9(10.5)$~\cite{Akdag:2023oob}  & $(210 \  |  \ 26) $ & $( 0.10 \ | \ 0.11) $\\
$g_{\omega\eta\eta'}~~[\text{GeV}^{-2}] $  & 0.43(0.69) \,
Eq.~\eqref{eq:upper_limit_vector_coupling_Eta'}
 & $(130 \   | \ 17)$ & $( 0.062\ | \  0.069)$\\
\bottomrule
\end{tabular}
\renewcommand{\arraystretch}{1.0}
\caption{Empirical values, upper limits, and  lower thresholds (sensitivities) of the BSM  hadronic coupling constants $g_1$, $g_1\delta g_1$, and $g_{\omega\eta\eta'}$, derived  (i) from the $\eta' \to \eta\pi^+\pi^-$ Dalitz-plot asymmetries of 
Ref.~\cite{Akdag:2023oob} or computed from Eq.~\eqref{eq:upper_limit_vector_coupling_Eta'},
(ii) from the experimental upper bounds on the  branching ratios $\B_{\eta'\to\eta \ell^+ \ell^-}$ given in
Table~\ref{tab:BR+SM} for the  ($e^+e^-$ | $\mu^+\mu^-$)
dilepton pairs, as calculated using the relations~\eqref{eq:WidthsEtapEtacoeff}
assuming that the other two coupling constants are vanishingly small, and (iii) as in (ii), but from the SM predictions listed in Table~\ref{tab:BR+SM}. 
\label{tab:BR_etap}}
\end{table}

Assuming that the other two coupling constants are negligible, the relations~\eqref{eq:WidthsEtapEtacoeff} can also be used to derive upper bounds for one of the three BSM coupling constants at a time, where the product $g_1\delta g_1$ can be non-zero even if $g_1$ itself is negligible.
The experimental limits of the branching ratios and the Standard-Model predictions for the dilepton branching ratios based on the $C$-even two-photon mechanism~\cite{Schafer:2023qtl} are taken from Table~\ref{tab:BR+SM}. Again
the latter are used to calculate the lower threshold for the above-mentioned coupling constants at which the BSM scenario would no longer dominate the Standard Model. The upper bounds
and the lower thresholds (sensitivities) are listed in the third and fourth columns of the Table~\ref{tab:BR_etap}, respectively. 
In the second column of Table~\ref{tab:BR_etap}, we list the empirical values of $g_1$, of the product $g_1\delta g_1$, both derived via the $\eta'\to \eta\pi^+\pi^-$ Dalitz-plot asymmetries in Ref.~\cite{Akdag:2023oob}, and  of $g_{\omega\eta\eta'}$, indirectly calculated via Eq.~\eqref{eq:g_rho-pi-eta} and 
Eq.~\eqref{eq:upper_limit_vector_coupling_Eta'}
from the corresponding $\eta\to\pi^0\pi^+\pi^-$ Dalitz-plot asymmetries~\cite{Akdag:2021efj,Akdag:2023oob}.

The empirical path via the
$\eta'\to \eta\pi^+\pi^-$ Dalitz-plot asymmetries and Eq.~\eqref{eq:upper_limit_vector_coupling_Eta'} to determine the BSM coupling constants is clearly preferable to the upper limit from  the experimental bounds on the $\eta\to \pi^0\ell^+\ell^-$ decays, but still compatible with the lower thresholds from the  $C$-even two-photon process of the Standard Model. 
However, the gaps between the empirical values and  lower thresholds are smaller than in the $\eta$ cases, especially for
$g_1$.

\section{Summary and outlook}
\label{sec:summary}
In this work, we have studied the $C$- and $CP$-violating decays $\eta\to\pi^0 \ell^+ \ell^-$ and $\eta'\to\eta \ell^+ \ell^-$, which can---from a phenomenological point of view---be driven by three different mechanisms. The first two of these are short-distance contributions induced by semi-leptonic four-point vertices and long-distance contributions caused by $C$- and $CP$-odd photon--hadron couplings. The only statements we can make about them is that they contribute as constants to respective transition form factors at leading order in ToPe$\chi$PT~\cite{Akdag:2022sbn}, that they cannot be distinguished by a sole measurement of the semi-leptonic decay widths, and that NDA estimates them to be of the same order of magnitude. 

In contrast, the third mechanism, i.e., long-distance contributions induced by hadronic intermediate states, is conceptually more insightful. To access these contributions we have established dispersion relations for the isovector contribution to the transition form factors $\eta(\to\pi^0\pi^+\pi^-)\to\pi^0\gamma^*$ and $\eta'(\to\eta\pi^+\pi^-)\to\eta\gamma^*$. 
By construction, these form factors meet the fundamental requirements of analyticity and unitarity, solely relying on the dominant hadronic contribution of the $P$-waves in the $C$- and $CP$-odd $\eta\to\pi^0\pi^+\pi^-$ and $\eta'\to\eta\pi^+\pi^-$ amplitudes, which have been worked out in Refs.~\cite{Akdag:2021efj,Akdag:2023oob}. The non-perturbative predictions thereby obtained allow us to directly investigate the correlation between $C$-violating signals in different decays in a model-independent manner. By an analytic continuation of the $C$-odd $\eta\to\pi^0\pi^+\pi^-$ and $\eta'\to\eta\pi^+\pi^-$ $P$-wave amplitudes to the second Riemann sheet, we have extracted $C$-odd $\rho$-meson couplings to $\eta\pi^0$ and $\eta'\eta$.  These two couplings can be related by total isospin and NDA to coupling constants entering the isoscalar contribution in a VMD model for $\eta'\to\eta\omega\to\eta\gamma^*$ and $\eta\to\pi^0\omega\to\pi^0\gamma^*$, respectively. However, the $\omega\to\eta\pi^0$ coupling could 
be bounded more precisely from the empirical limit on the corresponding partial width.

Accounting for these hadronic long-range effects only, we have predicted the corresponding upper limits on the semi-leptonic decay widths, relying on ToPe forces in the respective purely hadronic three-body decays as input. 
We observed that the currently most precise measurements of $\eta\to\pi^0 \ell^+ \ell^-$ have a similar sensitivity to isoscalar ToPe interactions as the measured Dalitz-plot asymmetries in $\eta\to\pi^0\pi^+\pi^-$, despite their different scaling with small BSM couplings. 
As we found the experimental limits for $\eta\to\pi^0 \ell^+ \ell^-$ to be more restrictive than our theoretically predicted ones, we were able to use the respective transition form factor as a constraint to sharpen the bounds on $C$~violation in $\eta\to\pi^0\pi^+\pi^-$.  On the other hand, both the isotensor coupling for $C$-odd $\eta$ decays and the corresponding $\eta'\to\eta \ell^+ \ell^-$ effects are more rigorously constrained indirectly from Dalitz-plot asymmetries.  
Finally, we have determined the best possible sensitivity to all $C$-odd hadronic couplings in semi-leptonic decays due to the Standard-Model background.

Given the relatively loose experimental bounds, we have largely eschewed concrete estimates of \emph{theoretical} uncertainties in our limit setting.  The dispersion relation used to connect hadronic and semi-leptonic decays are expected to work extremely well in the elastic approximation, with two-pion intermediate states only, for the relevant vector-isovector channel: comparable sum rules for SM processes typically work to better than 10\%~\cite{Schneider:2012ez,Hanhart:2016pcd}.  This is the relevant accuracy for the relation between both types of $C$-odd effects.  Neglected higher-order corrections in the chiral or large-$N_c$ expansions, which may easily amount to 30\% or so, only affect the interpretation of effective coupling constants on the mesonic level in terms of underlying LEFT or SMEFT operators~\cite{Akdag:2022sbn}.

Further perspectives on the decays $\eta\to\pi^0 \ell^+ \ell^-$ and $\eta'\to\eta \ell^+ \ell^-$ could be opened by possible future measurements of the respective Dalitz-plot distributions.  
This would allow us to investigate actual $C$- and $CP$-odd observables, the Dalitz-plot asymmetries arising from the interference with the respective SM contributions.   Such interference effects would, as the asymmetries in the hadronic $\eta$ and $\eta'$ decays, scale linearly with BSM couplings, however with likely less advantage in sensitivity due to the strongly suppressed SM amplitudes.
Significant asymmetries can only be expected if, accidentally, $C$-even and -odd amplitudes happen to be of similar size.
Still, due to synergy effects with other BSM searches in these decays, such as for weakly coupled light scalars~\cite{Gan:2020aco}, renewed experimental efforts are strongly encouraged.

\acknowledgments
We would like to thank Danny van Dyk for useful discussions.
Financial support by the Avicenna-Studienwerk e.V.\ with funds from the BMBF,
as well as by the DFG (CRC 110, ``Symmetries and the Emergence of Structure in QCD''),
is gratefully acknowledged.

\bibliographystyle{JHEP_mod}
\bibliography{base}

\end{document}